\documentclass{ws-procs975x65}

\usepackage{amssymb,amsmath}

\newcommand{\mbf}[1]{\boldsymbol{#1}}

\newcommand{\eq}{\begin{equation}}
\newcommand{\eqend}{\end{equation}}
\newcommand{\eqa}{\begin{eqnarray}}
\newcommand{\nonueqa}{\begin{eqnarray*}}
\newcommand{\eqaend}{\end{eqnarray}}
\newcommand{\nonueqaend}{\end{eqnarray*}}

\newcommand{\bma}[1]{\begin{array}{#1}}
\newcommand{\ema}{\end{array}}
\newcommand{\bc}{\begin{center}}
\newcommand{\ec}{\end{center}}

\renewcommand{\theequation}{\thesection.\arabic{equation}}

\newcommand{\complex}{{\mathbb C}} 
\newcommand{\zed}{{\mathbb Z}} 
\newcommand{\nat}{{\mathbb N}} 
\newcommand{\real}{{\mathbb R}} 
\newcommand{\sphere}{{\mathbb S}} 
\newcommand{\rat}{{\mathbb Q}} 
\newcommand{\id}{{1\!\!1}} 

\newif\ifold             \oldtrue

\def\nn{\nonumber}
\def\bz{{\overline{z}}}

\newcommand{\Tr}[1]{\:{\rm Tr}\,#1}

\def\e{{\,\rm e}\,}

\def\be{\begin{equation}}
\def\ee{\end{equation}}
\def\bea{\begin{eqnarray}}
\def\eea{\end{eqnarray}}
\def\bd{\begin{displaymath}}
\def\ed{\end{displaymath}}

\def\dd{{\rm d}}

\def\ii{{\,{\rm i}\,}}

\newcommand{\beq}{\begin{eqnarray}}
\newcommand{\eeq}{\end{eqnarray}}

\newcommand{\DD}{{\rm D}}

\makeatletter
\newdimen\normalarrayskip              
\newdimen\minarrayskip                 
\normalarrayskip\baselineskip
\minarrayskip\jot
\newif\ifold             \oldtrue            
\def\arraymode{\ifold\relax\else\displaystyle\fi} 
\def\@arrayskip{\ifold\baselineskip\z@\lineskip\z@
     \else
     \baselineskip\minarrayskip\lineskip2\minarrayskip\fi}
\def\@arrayclassz{\ifcase \@lastchclass \@acolampacol \or
\@ampacol \or \or \or \@addamp \or
   \@acolampacol \or \@firstampfalse \@acol \fi
\edef\@preamble{\@preamble
  \ifcase \@chnum
     \hfil$\relax\arraymode\@sharp$\hfil
     \or $\relax\arraymode\@sharp$\hfil
     \or \hfil$\relax\arraymode\@sharp$\fi}}
\def\@array[#1]#2{\setbox\@arstrutbox=\hbox{\vrule
     height\arraystretch \ht\strutbox
     depth\arraystretch \dp\strutbox
     width\z@}\@mkpream{#2}\edef\@preamble{\halign \noexpand\@halignto
\bgroup \tabskip\z@ \@arstrut \@preamble \tabskip\z@ \cr}%
\let\@startpbox\@@startpbox \let\@endpbox\@@endpbox
  \if #1t\vtop \else \if#1b\vbox \else \vcenter \fi\fi
  \bgroup \let\par\relax
  \let\@sharp##\let\protect\relax
  \@arrayskip\@preamble}
\makeatother

\begin{document}

\title{Strings, Gauge Fields and Membranes\footnote{\uppercase{T}o be
  published in the
  \uppercase{I}an \uppercase{K}ogan \uppercase{M}emorial
  \uppercase{C}ollection ``\uppercase{F}rom \uppercase{F}ields to
  \uppercase{S}trings: \uppercase{C}ircumnavigating
  \uppercase{T}heoretical \uppercase{P}hysics'', \uppercase{W}orld
  \uppercase{S}cientific, 2004.}}

\author{RICHARD J. SZABO}

\address{Department of Mathematics\\ Heriot-Watt University\\
Scott Russell Building\\ Riccarton, Edinburgh EH14 4AS, U.K.\\
Email: R.J.Szabo@ma.hw.ac.uk}

\maketitle

\abstracts{
We present an overview of the intimate relationship between
string and D-brane dynamics, and the dynamics of gauge and gravitational
fields in three spacetime dimensions. The successes, prospects and
open problems in describing both perturbative and nonperturbative aspects
of string theory in terms of three-dimensional quantum field theory
are highlighted.}

\centerline{\small{\tt HWM-04-9 , EMPG-04-05 , hep-th/0405289}}
\centerline{\small May 2004}

\vspace{0.5cm}
\tableofcontents
\newpage

\section{Introduction\label{Intro}}
\setcounter{equation}{0}

In this article we will describe a very simple example of a holographic
correspondence, that between two-dimensional conformal field theories
and three-dimensional topological quantum field theories. Generally,
this duality is described as an isomorphism between the space of
conformal blocks on
the conformal field theory side and the space of physical states on
the topological field theory side (where by ``physical'' we mean, for
instance, gauge invariant). The goal is to compute correlators of
two-dimensional conformal field theory, with open or closed
worldsheets, from such ``bulk'' three-dimensional quantum field
theories. The best known example of this holography is the equivalence
between a Chern-Simons gauge theory defined on a three-manifold $M_3$
with boundary and a Wess-Zumino-Novikov-Witten (WZNW) model on
$\partial M_3$~\cite{Witten1,MS1}.

On quite general grounds, any conformal field theory can be shown to
give rise to a topological quantum field theory by extracting a modular tensor
category from the conformal field theory chiral vertex operator
algebra~\cite{FRS1}. For example, the Moore-Seiberg data of a rational
conformal field theory
encode the basic braiding, fusing and $\sf S$-matrices, along with the
appropriate pentagon and hexagon identities~\cite{MS2}. They give rise to a
topological quantum field theory in three-dimensions which can be used
to compute invariants of knots and links in three-manifolds. They also
give rise to a modular tensor category $\mathcal{C}$, the category of
representations of the rational chiral vertex operator algebra, which
may be thought of as a basis-independent formulation of the
Moore-Seiberg data. $\mathcal{C}$ generalizes the well-known category of
finite-dimensional vector spaces. With this one can develop a powerful
graphical calculus in terms of ribbon graphs which correspond to
framed Wilson lines in three-dimensions~\cite{FFFS1}.

One problem with this correspondence is that the Hilbert space
$\mathcal{H}$ of a topological field theory is only isomorphic to the
space of {\it holomorphic} conformal blocks of the associated
conformal field theory. A chiral correlator in the conformal field
theory is completely determined by a choice of vector
in~$\mathcal{H}$. Thus the equivalence just described is not exactly an
example of a {\it holographic} correspondence, in which the full
conformal field theory correlation functions, comprising both
holomorphic and antiholomorphic sectors, on the boundary would be
reproduced by some three-dimensional theory in the bulk. The goal is
then to find some three-dimensional theory that corresponds to the
{\it full} conformal field theory.

Once this correspondence is achieved, we can try to use it to
reformulate perturbative string theory in the simpler language of quantum
field theory. But in order to implement the various nonperturbative
ingredients of string theory, we need to somehow add additional
structure to this formalism. In this article we shall describe how to do
this in the simplest instance of WZNW models, describing string
propagation in group manifolds, which can always be constructed by
using Chern-Simons theories~\cite{MS1}. We will thereby derive string
theory from a theory of higher-dimensional extended objects which we might dub
``Worldsheet M-Theory'', because all string theories will originate
from a single bulk theory in three dimensions.

\subsection{Topological Membranes\label{TM}}

\begin{figure}[ht]
\centerline{\epsfxsize=5cm\epsfbox{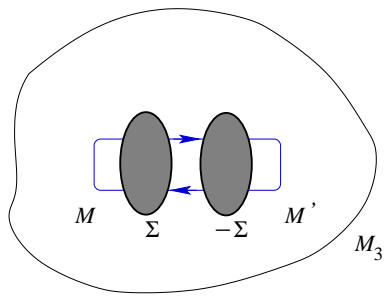}}
\caption{A Heegaard splitting of a three-manifold $M_3$. The
  three-manifold is cut along a Riemann surface $\Sigma$ into two
  three-manifolds $M$ and $M'$, whose boundaries are identified by an
  orientation reversing homeomorphism as $\Sigma=\partial
  M$, $-\Sigma=\partial M'$. The closed curve represents a Wilson loop in the
  bulk. \label{Heegaard}}
\end{figure}

\begin{figure}[ht]
\centerline{\epsfxsize=2cm
\epsfbox{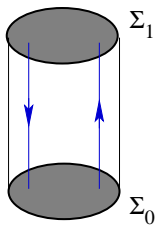}}
\caption{The three-manifold $M_3=\Sigma\times[0,1]$. The Wilson lines
  propagate between the two boundaries $\Sigma_0$ and
  $\Sigma_1$, which are each copies of $\Sigma$ but carry opposite
  orientations.}
\label{Sigmax01}\end{figure}

The higher-dimensional object mentioned above will be refered to as a
``topological membrane''~\cite{Kogan1,CK1}, and it is obtained by
filling in the string
worldsheet and viewing it as the boundary of a three-manifold. Despite
some similarities which we describe below, these membranes are
fundamentally different from the membrane degrees of freedom in
11-dimensional M-Theory. There are various ways to regard the induced
string theory from three-dimensions. For instance, we may identify the
string worldsheet $\Sigma$ with the boundary of the three-manifold
$M_3$ in question, $\Sigma=\partial M_3$, but as pointed out above the
induced field theory carries only chiral degrees of freedom. Instead,
we shall proceed with the observation~\cite{Witten1} that locally the
three-manifold can always be subjected to a Heegaard splitting
$M_3=M\#_\Sigma M'$, as depicted in Figure~\ref{Heegaard}, and the
holomorphic and antiholomorphic sectors of the two-dimensional theory can be
identified with the induced degrees of freedom on the two boundaries
$\partial M$ and $\partial M'$. In a neighbourhood of this slice,
$M_3$ may be identified with the three-manifold $\Sigma\times[0,1]$
depicted in Figure~\ref{Sigmax01}. The two boundary components
$\Sigma_0=\Sigma\times\{0\}$ and $\Sigma_1=\Sigma\times\{1\}$ then
respectively contain the holomorphic and antiholomorphic degrees of
freedom of the induced two-dimensional conformal field
theory~\cite{EMSS1}. This is the picture that we shall always have in
mind throughout all our string constructions.

A crucial point will be that in order to see this program through it
will be necessary to destroy the topological invariance of the
original bulk quantum field theory. The dynamical ingredients which
will go into the construction of the topological membrane can be summarized as
follows. In addition to the Chern-Simons term, we will add the
Yang-Mills action and consider a topologically massive gauge theory in
the bulk~\cite{EMSS1}--\cite{DJT1}. The presence of propagating gluon
degrees of freedom will
enable us to control boundary conditions simultaneously on the
left-moving and right-moving worldsheets $\Sigma_0$ and $\Sigma_1$,
and to thereby induce the full non-chiral conformal field theory on
$\Sigma$~\cite{Cooper1}. We will then couple this gauge theory to topologically
massive gravity~\cite{DJT1}, consisting of the Einstein-Hilbert and
gravitational Chern-Simons actions in three dimensions, which will
have the effect of inducing two-dimensional quantum gravity on the
boundary $\Sigma$~\cite{Carlip1,Kogan2}. A further conformal coupling
to a three-dimensional scalar field theory will then produce the
dilaton field~\cite{KMS1}, and hence the string
coupling~$g_s$. Finally, minimal couplings of the gauge theory to
charged matter fields in the bulk will yield deformations of the induced
conformal field theory~\cite{Kogan3} enabling us to construct vertex
operators and, ultimately, states corresponding to D-branes~\cite{bb}.

Let us conclude this introduction with some of the primary motivations
for rewriting string theory this way in terms of topological membranes:
\begin{enumerate}
\item Many aspects of string dynamics have natural interpretations in
  terms of the dynamics of gauge and gravitational fields in the
  bulk.
\item Various algebraic properties of two-dimensional conformal field
  theories can be understood geometrically and dynamically in the
  three-dimensional picture. In this sense, important dynamical
  effects are responsible for fundamental properties of the induced
  conformal field theory. This yields new dynamical perspectives on
  string constructions, which are simpler in the language of
  three-dimensional quantum field theories.
\item As a ``Worldsheet M-Theory'', it has the highly desirable
  feature that the basic principles inherent to three-dimensional
  gauge theory are far fewer than those to all of the existing string
  theories.
\item There are various pieces of evidence that the topological
  membrane framework gives a potential dynamical origin for the
  eleventh extra dimension of M-Theory from fundamental string
  fields. First of all, the conformal scalar coupling mentioned above
  can be adjusted to induce a dynamical string coupling $g_s$~\cite{KMS1},
  analogously to the way that the string coupling emerges as a
  dynamical variable in M-Theory~\cite{Witten2}. Secondly, the
  topological membrane construction is precisely a lower-dimensional
  version of the Ho\v{r}ava-Witten mechanism~\cite{HW1,HW2}, in which
  the cancellation of gauge and
  gravitational anomalies on the boundaries of an open cylindrical
  membrane stretched between the two boundaries of an 11-dimensional
  spacetime $\real^{10}\times\sphere^1/\zed_2$ requires there to be
  one $E_8$ gauge group on each spacetime boundary, and it thereby
  induces the $E_8\times E_8$ heterotic string theory. However, this
  latter approach relies crucially on spacetime structures such as
  11-dimensional supergravity, whereas in the present case the
  emphasis is on the worldsheet properties of the open topological
  membrane. Finally, the supermembrane in the spectrum of
  11-dimensional supergravity and the topological membrane, although
  not identical, do share some amusing similarities. The supermembrane
  couples to a three-form supergravity potential whose action is the
  11-dimensional Chern-Simons term~\cite{CJS1}, while its worldvolume
  description when it wraps degenerate three-cycles of a Calabi-Yau
  manifold is a three-dimensional Chern-Simons
  theory~\cite{OLoughlin1}. All of these similarities hint that the
  extra dimension of topological membrane theory could potentially be
  embedded in 11-dimensional M-Theory.
\end{enumerate}

\section{Chern-Simons Gauge Theory\label{CSGT}}
\setcounter{equation}{0}

We will begin with an investigation of pure Chern-Simons gauge theory
and hence describe the basic holographic duality with chiral conformal
field theories.

\subsection{The Chern-Simons Action\label{TCSA}}

Let $M_3$ be a compact oriented three-manifold, possibly with
boundary, and let $G$ be a compact connected Lie group equipped with an
invariant bilinear form $\Tr$ on its Lie algebra. The Chern-Simons
action is defined by the functional~\cite{Witten1,EMSS1,CS1,Schwarz1}
\bea
S_{\rm CS}[A]&=&\frac k{4\pi}\,\int_{M_3}\Tr\left(A\wedge\dd A+
\mbox{$\frac23$}\,A\wedge A\wedge A\right)\nn\\&&+\,\frac k{4\pi}\,
\int_{\partial M_3}\Tr\left(A^{1,0}\wedge A^{0,1}\right)
\label{SCSAdef}\eea
where $A$ is a connection one-form on a principal $G$-bundle $P\to
M_3$. The first term defines a three-dimensional topological quantum
field theory, containing no gluons or explicit dependence on any
metric of $M_3$. The second term ensures that the action has classical
extrema when $M_3$ has a non-empty boundary and it depends on the
choice of a complex structure on $\partial M_3$. $A^{1,0}$
(resp. $A^{0,1}$) denotes the holomorphic (resp. antiholomorphic)
component of the gauge connection on $\partial M_3$ with respect to
this auxilliary complex structure.

When the gauge group $G$ is simple, it has non-trivial homotopy group
$\pi_3(G)=\zed$. Then the action (\ref{SCSAdef}) is not invariant
under homotopically non-trivial (large) gauge transformations but
instead changes by $2\pi\,k\,w$, where $w\in\zed$ is the winding
number of the map $M_3\to G$~\cite{DJT1}. Consistency of the quantum
field theory, in a path integral formalism to be described below, then
requires the quantization condition $k\in\zed$ on the Chern-Simons
coefficient in (\ref{SCSAdef}).

\subsection{Path Integral Formulation\label{PIF}}

We will study the Chern-Simons path integral with prescribed boundary
conditions for the gauge field $A$ on the Riemann surface
$\Sigma=\partial M_3$, and hence define the Hartle-Hawking type state
\beq
\mathcal{F}[\mbf A]=\int_{A^{1,0}=\mbf A}\DD A~\e^{\ii S_{\rm CS}[A]}
\ .
\label{HHtypestate}\eeq
This state defines a vector in the Chern-Simons Hilbert space
$\mathcal{H}_\Sigma^{\rm CS}$ which depends only on the topology and
framing of the three-manifold $M_3$~\cite{Witten1}. It determines a
modular functor ${\mathcal F}:\mathcal{H}_\Sigma^{\rm CS}\to\complex$
defined by
\beq
\mathcal{F}(\Psi)=\langle\mathcal{F}|\Psi\rangle=
\frac1{{\rm vol}\,\mathcal{G}_\Sigma}\,\int\DD\mbf A~
\overline{\DD\mbf A}~\overline{\mathcal{F}[\mbf A]}\,\Psi[\mbf A] \ ,
\label{modfunctordef}\eeq
where $\mathcal{G}_\Sigma=C^\infty(\Sigma,G)$ is the group of gauge
transformations on the boundary $\Sigma$. The formula
(\ref{modfunctordef}) utilizes the natural inner product on the Hilbert
space $\mathcal{H}_\Sigma^{\rm CS}$ usually defined in holomorphic
(coherent state) quantization~\cite{BN1,LR1}.

Given a connection one-form $A$, let us now parametrize its gauge
orbit as
\beq
A=g^{-1}\,\bar A\,g+g^{-1}~\dd g \ ,
\label{gaugeorbit}\eeq
where $g:\Sigma\to G$ is a gauge transformation on the boundary and
$\bar A$ is a fixed representative of the gauge equivalence class on
$M_3$ of the gauge field $A$. With standard gauge-fixing techniques,
one then finds that the path integral (\ref{HHtypestate}) factorizes
as~\cite{MS1,Ogura1}
\beq
\mathcal{F}[\mbf A]=\int_{\bar A^{1,0}=\mbf A}\DD\bar A~\delta\left(
F_{\bar A}\right)\,\Delta_{\rm FP}\left(\bar A\,\right)~
\e^{\ii S_{\rm CS}[\bar A\,]}\,\int\DD g~
\e^{\ii k\,S_{\rm W}^+[g,\mbf A]} \ ,
\label{HHstatefact}\eeq
where $F_A=\dd A+A\wedge A$ is the curvature of $A$, $\Delta_{\rm FP}$
denotes the usual Faddeev-Popov determinant, and the delta-function in
(\ref{HHstatefact}) restricts the functional integration to classical gauge
field configurations (with the specified boundary conditions on
$\Sigma$) which extremize the action (\ref{SCSAdef}). The functional
\bea
S_{\rm W}^+[g,\mbf
A]&=&\frac1{4\pi}\,\int_\Sigma\Tr\left(\left|g^{-1}\,\partial g
\right|^2-2g^{-1}\,\overline{\partial}g\wedge\mbf A\right)\nn\\&&
+\,\frac1{12\pi}\,\int_{M_3}\Tr\left(g^{-1}~\dd g\right)^3
\label{SWZNWchiral}\eea
is the action of the chiral gauged WZNW model for the group field $g$
coupled to an external gauge field $\mbf A$ on the Riemann surface
$\Sigma=\partial M_3$~\cite{MS1,EMSS1,Witten3}.

The induced boundary conformal field theory is
chiral precisely because of the background field coupling, which
leaves only one conserved current $g^{-1}\,\partial g$. After setting
$\mbf A=0$ by an appropriate choice of boundary conditions, the
Chern-Simons path integral exhibits a complete decoupling into bulk
and surface degrees of
freedom. In other words, on a three-manifold with boundary the gauge
symmetry of the field theory defined by the action (\ref{SCSAdef})
becomes anomalous, and we absorb this anomaly by identifying new
degrees of freedom on the boundary.

\subsection{Phase Space\label{PS}}

Near the boundary $\partial M_3$, the three-manifold may be regarded
topologically as the product $M_3=\Sigma\times\real$ and we may
analyse the Chern-Simons gauge theory using a standard Hamiltonian
formalism~\cite{Witten1}. The canonical variables are the two
components of the gauge
field $\mbf A$ on $\Sigma$, and the Gauss law of (\ref{SCSAdef}) shows that the
phase space $\mathcal{P}_\Sigma^{\rm CS}$ of Chern-Simons theory is
the moduli space of flat $G$-connections, $F_{\mbf A}=0$, on $\Sigma$ modulo
gauge transformations. If $\Sigma$ is a compact oriented Riemann surface of
genus $g\geq0$ with $n>2-2g$ punctures, then the phase space has
the explicit presentation
\beq
\mathcal{P}_\Sigma^{\rm CS}~=~{\rm Hom}\bigl(\pi_1(\Sigma)\,,\,G
\bigr)~/~G
\label{phasespacedef}\eeq
reflecting the fact that flat connections are determined entirely by
their holonomies (Wilson lines) around loops on $\Sigma$. This is a
symplectic manifold of finite dimension
\beq
\dim\mathcal{P}_\Sigma^{\rm CS}=(2g+n-2)\,\dim G \ ,
\label{dimPCS}\eeq
with symplectic leaves obtained by restricting the holonomies of
$\mbf A$ around the punctures to lie in conjugacy classes of the
gauge group $G$. As the volume of the leaves is finite, quantization
of (\ref{phasespacedef}) will produce a finite dimensional Hilbert
space $\mathcal{H}_\Sigma^{\rm CS}$. As we will discuss in more detail
later on, with appropriate parabolic conditions at the punctures in
the path integral formalism, $\mathcal{H}_\Sigma^{\rm CS}$ coincides
with the finite dimensional vector space of conformal blocks of a
chiral WZNW model on a genus $g$ surface $\Sigma$ with $n$ primary
vertex operator insertions~\cite{Witten1}. The dimension of this space
is computed by the Verlinde formula~\cite{Verlinde1}.

\subsection{Conformal Field Theory Constructions\label{CFTC}}

Let us now summarize the list of (chiral) conformal field theories
that can be constructed in this way from pure Chern-Simons gauge
theory:
\begin{enumerate}
\item One can build a multitude of rational conformal field theories
in the Chern-Simons formalism using GKO coset constructions~\cite{GKO1}. Let
$H\subset G$ be a subgroup such that the Virasoro algebra over $G$ can
be decomposed into the orthogonal direct sum of the Virasoro algebra
over $H$ and the Virasoro algebra over the coset $G/H$. Then one can
construct the coset current algebra based on $G_k/H_l$ with action
$k\,S_{\rm W}^{+}[g]-l\,S_{\rm W}^{+}[h]$ for group fields
$g:\Sigma\to G$ and $h:\Sigma\to H$. This theory is holographically
dual to three-dimensional Chern-Simons gauge theory with action
$k\,S_{\rm CS}[A]-l\,S_{\rm CS}[B]$, where $A$ is a $G$-connection and
$B$ an $H$-connection on $M_3$~\cite{MS1}. For example, the standard minimal
models can be obtained through the coset constructions
\beq
\mathcal{M}_k~=~SU(2)_k\times SU(2)_1~/~SU(2)_{k+1}
\label{minmodels}\eeq
where the quotient is by the diagonal $SU(2)$ action. In this way it
is possible to characterize the zoo of rational conformal field
theories on the basis of a single gauge theory in three-dimensions.
\item Supersymmetric extensions of all of these constructions are
also possible and straightforward to carry out, giving the standard
$N=1$~\cite{ST1} and $N=2$~\cite{CKS1,CKS2} superconformal field
theories in two dimensions. They will not be discussed in this
article.
\item The holographic correspondence can also be extended to more
  general conformal field theories using model-independent topological
  field theory formulations~\cite{FRS1,FFFS1}. These more general scenarios are
  axiomatic and no action formalism is possible for them in the manner
  described above. They also will not be dealt with in this article.
\end{enumerate}

\section{Topologically Massive Gauge Theory\label{TMGT}}
\setcounter{equation}{0}

We will now modify the Chern-Simons action (\ref{SCSAdef}) by adding
to it a Yang-Mills term. If it is not present initially, then it will
be induced in any case by quantum radiative corrections when the gauge
theory is minimally coupled to charged matter, as we will do later on. As we
explain, the inclusion of this term enables us to significantly expand
the list of conformal field theory constructions of the previous section and
to begin working our way towards building up complete string theories.

\subsection{Classical Aspects\label{CA}}

Let us fix a Lorentzian metric on $M_3$ of signature $(2,1)$ and
denote the corresponding Hodge duality operator by
$*:\Omega^p(M_3,{\rm ad}\,P)\to\Omega^{3-p}(M_3,{\rm ad}\,P)$. The
action of topologically massive gauge theory is defined by the
functional~\cite{Schonfeld1,DJT1}
\bea
S_{\rm TMGT}[A]&=&\int_{M_3}\Tr\left[-\frac1{e^2}\,F_A\wedge*F_A+\frac
k{4\pi}\,A\wedge\left(\dd A+\mbox{$\frac23$}\,A\wedge A\right)\right]
\nn\\&&+\,\int_{\partial M_3}\Tr\left(A^{1,0}\wedge\Pi^{0,1}\right) \
,
\label{STMGTAdef}\eea
where the Yang-Mills coupling constant $e^2$ has dimensions of mass
and
\beq
\Pi=\frac k{4\pi}\,A-\frac2{e^2}\,*F_A
\label{Pi01def}\eeq
is the canonical momentum conjugate to the gauge field $A$. The second
term in (\ref{STMGTAdef}) imposes conformal Dirichlet boundary
conditions fixing the connection components $A^{1,0}$ on
$\Sigma=\partial M_3$.

The Yang-Mills term in (\ref{STMGTAdef}) is gauge invariant even when
$\partial M_3\neq\emptyset$, and so the same chiral WZNW model as in
the last section is induced on the boundary within the path integral
framework for the action (\ref{STMGTAdef}). The crucial consequence of
its addition is that the bulk field theory is no longer topological, because it
depends explicitly on the metric of $M_3$ through the Hodge
operator. In contrast to the pure Chern-Simons gauge theory, there are
now propagating gluon degrees of freedom and, as we show in the next
subsection, the Hilbert space of physical states becomes
infinite-dimensional. To see this, we let $D_A=\dd+A$ denote the usual
gauge covariant derivative, and note that one can write the
Euler-Lagrange equations arising from the action (\ref{STMGTAdef}) in
the form
\beq
\left(D_A^2-\mu^2\right)*F_A=*\left(*F_A\wedge*F_A\right) \ .
\label{TMGTEOM}\eeq
This is the equation of motion for a single propagating degree of freedom
with topological mass
\beq
\mu=\frac{e^2|k|}{4\pi} \ .
\label{topmassTMGT}\eeq

Because of the presence of a massive gluon, the Yang-Mills term serves
as an infrared regularization of the pure Chern-Simons theory, which
is recovered in the infrared limit $e^2\to\infty$ wherein the energies
of all states containing the gluon decouple from the rest of
the spectrum. We may thereby regard pure Chern-Simons theory as the
ground state of topologically massive gauge theory. A particularly
important consequence of the presence of this new degree of freedom is
the enlargement of the phase space of the gauge theory. In the pure
Chern-Simons case of the previous section, there are only two
canonical variables $A^{1,0}$ and $A^{0,1}$ which are conjugate to
each other, and so it is not possible to fix both components of the
gauge field at the same time on $\partial M_3$, as this
would violate the canonical Poisson brackets. In contrast, the action
(\ref{STMGTAdef}) involves {\it four} independent canonical variables
$\Pi^{1,0}$, $\Pi^{0,1}$, $A^{1,0}$ and $A^{0,1}$. Now $A^{1,0}$ and
$A^{0,1}$ Poisson commute, and it is consistent to fix both gauge field
components on the boundary.

Furthermore, one can now vary the choice of worldsheet complex
structure in the induced conformal field theory on $\Sigma$ via its
coupling to the metric of $M_3$~\cite{CK1}. This means that we can now
generate both holomorphic and
antiholomorphic degrees of freedom on the connected components of
$\partial M_3=\Sigma_0\amalg\Sigma_1$ holding different induced
conformal field theories, one in each boundary component as
illustrated in Figure~\ref{Sigmax01}. Alternatively, we may choose to
kill all degrees of freedom in one boundary component by fixing both
$A^{1,0}$ and $A^{0,1}$ there, leading us into various string
constructions~\cite{Cooper1}. For example, the heterotic string
construction starts with a topological membrane that has a semi-simple
gauge group $G_{\rm L}\times G_{\rm R}$, where $G_{\rm L}$ is the
gauge group for the
ordinary bosonic topologically massive gauge theory, and $G_{\rm R}$
is the gauge group for the $N=1$ supersymmetric extension of
topologically massive gauge theory~\cite{CK1,ST1}. The heterotic worldsheet is
constructed by placing conformal boundary conditions on $\Sigma_0$ and
killing all degrees of freedom on $\Sigma_1$ for the theory based on
$G_{\rm L}$, thereby inducing bosonic left-moving worldsheet degrees
of freedom, while killing all modes on $\Sigma_0$ and selecting
conformal boundary conditions on $\Sigma_1$ for the $G_{\rm R}$
theory, leading to supersymmetric right-moving degrees of freedom on
the string worldsheet. Thus the seemingly hybrid nature of the
heterotic string construction becomes more natural in the topological
membrane framework, wherein the different sectors have a common
geometric origin in the choice of membrane boundary conditions.

\subsection{Hilbert Space\label{HS}}

Let us now analyse in detail the structure of the Hilbert space of
physical states of topologically massive gauge theory, focusing for
simplicity on the abelian case with gauge group $G=U(1)$,
corresponding to the $c=1$ conformal field theory of a single
boson~\cite{BN1}. Consider a
canonical split of the gauge connection $A=\mbf A+A_0~\dd t$ on a
spacetime of the form $M_3=\Sigma\times\real$, where $\Sigma$ is a
compact oriented surface of genus $g$. The metric on $M_3$ is of the
form $\dd s_\Sigma^2-\dd t^2$ while its orientation is given by the
three-form $\dd{\rm vol}_\Sigma\wedge\dd t$. Varying the action
(\ref{STMGTAdef}) with respect to $A_0$ yields the Gauss law, which in
the $A_0=0$ gauge reads
\beq
\frac1{e^2}~\dd*_2\dot{\mbf A}-\frac k{4\pi}\,F_{\mbf A}=0
\label{Gausslawabelian}\eeq
where $*_2$ is the Hodge operator on $\Sigma$ and a dot denotes
differentiation with respect to the time coordinate $t\in\real$. At
each fixed time slice $t\in\real$, we write the one-form $\mbf
A\in\Omega^1(\Sigma)$ using its Hodge decomposition as
\beq
\mbf A=\dd\xi+*_2\,\dd\chi+\mbf a \ ,
\label{Hodgedecompabelian}\eeq
where $\xi$ and $\chi$ are scalar fields on $\Sigma$ while $\mbf
a=\Bigl(a_1^{(\alpha)},a_2^{(\alpha)}\Bigr)_{\alpha=1,\dots,g}\in
{\rm H}^1(\Sigma,\real)=\real^{2g}$ are the harmonic degrees of freedom of
the gauge field.

Substituting (\ref{Hodgedecompabelian}) into the abelian version of the
action (\ref{STMGTAdef}) and applying the Gauss law constraint
(\ref{Gausslawabelian}) shows that the resulting action after
diagonalization decouples into $g+1$ independent pieces as
\beq
S_{\rm TMGT}[\mbf A]=S_{\rm f}[\varphi]+\sum_{\alpha=1}^gS_{\rm L}
\left[a_1^{(\alpha)}\,,\,a_2^{(\alpha)}\right] \ .
\label{STMGTdecouple}\eeq
The first term is the free particle action for the non-local scalar
field $\varphi=\sqrt{\nabla^2/e^2}\,\chi$ of mass $\mu$ given
by~\cite{DJT1}
\beq
S_{\rm f}[\varphi]=\frac12\,\int_{M_3}\left(\dd\varphi\wedge*\,\dd\varphi-\mu^2
*\varphi^2\right) \ ,
\label{freepartaction}\eeq
describing the dynamics of the single propagating gluon mode. The
second set of $g$ terms are topological and consist of identical
copies of the quantum mechanical action~\cite{KM1}--\cite{Wen1}
\beq
S_{\rm L}[a_1,a_2]=\int\dd t~\left(\frac1{2e^2}\,\dot a_i^2-\frac
k{4\pi}\,\epsilon^{ij}\,a_i\,\dot a_j\right)
\label{Landauaction}\eeq
where $\epsilon^{12}=+\,1$. This is the Landau action describing
propagation of a charged particle of mass $m=1/e^2$ on the plane
$(a_1,a_2)$ in a uniform magnetic field $B=k/4\pi$.

After diagonalization, the quantum Hilbert space of topologically
massive gauge theory is therefore given by
\beq
\mathcal{H}_\Sigma^{\rm TMGT}=\mathcal{H}_{\rm f}[\varphi]\otimes
\left(\mathcal{H}_{\rm L}\right)^{\otimes g} \ ,
\label{HilbertTMGT}\eeq
where the first factor is the infinite dimensional Hilbert space of
the free massive scalar field $\varphi$, while the second factor is the
Hilbert space for $g$ copies of the Landau problem on the
plane. $\mathcal{H}_{\rm L}$ is thus composed of infinitely many
Landau levels, with the mass gap between consecutive levels being equal to
$\Delta=|B|/m=\mu$, the mass of the gauge boson. This calculation ignores
large gauge transformations, which are parametrized by elements of the
lattice ${\rm H}^1(\Sigma,\zed)$ of rank $2g$. Demanding invariance of the
quantum gauge theory under them restricts the harmonic modes to the
torus $\mbf a\in {\rm H}^1(\Sigma,\real)/{\rm H}^1(\Sigma,\zed)$ and yields the
Hilbert space for the Landau problem on $\Sigma$. The Hilbert space
(\ref{HilbertTMGT}) then contains a dependence on the moduli of the Riemann
surface $\Sigma$ arising from the non-topological nature of the gauge
theory, but the mass gap $\Delta$ is always independent of these
moduli~\cite{Kogan5}. From this result we may also extract the pure
Chern-Simons Hilbert space $\mathcal{H}_\Sigma^{\rm CS}$ as the
projection of (\ref{HilbertTMGT}) onto the lowest Landau level in the
infrared limit~$\mu\to\infty$ (equivalently the strong coupling limit
$e^2\to\infty$).

\section{Hamiltonian Quantization\label{HQ}}
\setcounter{equation}{0}

In the previous sections we have used path integral quantization to
describe how boundary degrees of freedom are induced in order to
cancel the gauge anomaly of topologically massive gauge theory on a
three-manifold with boundary. Drawing from the
analysis of Section~\ref{HS}, we shall now investigate in detail how
these modes alternatively appear within the formalism of canonical
quantization. This will provide the basic building blocks for the
construction of quantum states of the topological membrane, which will
in turn be used to construct string amplitudes and to describe
non-perturbative string excitations of the membrane.

\subsection{Hamiltonian Formalism for Topologically Massive\\ Gauge
  Theory\label{HFTMGT}}

Let us consider a canonical split $A=\mbf A+A_0~\dd t=A_i~\dd
x^i+A_0~\dd t$ of the generic (non-abelian) topologically massive
gauge theory (\ref{STMGTAdef}) on the three-manifold
$M_3=\Sigma\times\real$. We define the electric and magnetic fields by
\bea
E&=&\Pi-\frac k{8\pi}\,*_2\mbf A \
, \label{electricfield}\\B&=&*_2F_{\mbf A} \ .
\label{magfield}\eea
The electric field (\ref{electricfield}) is analogous to the velocity
operator in the Landau problem. The Hamiltonian of topologically
massive gauge theory can then be written as
\beq
H_{\rm TMGT}=\int_\Sigma\Tr\left(e^2\,E\wedge*_2E+\frac1{e^2}\,
B\wedge*_2B\right) \ ,
\label{HTMGTdef}\eeq
while the Gauss law reads
\beq
*_2\mathcal{G}=\dd\,*_2E-\ii\mbf A\wedge E+\frac k{4\pi}\,*_2B\sim0
\ .
\label{GausslawTMGT}\eeq
The scalar field $\mathcal{G}$ on $\Sigma$ is the generator of
time-independent gauge transformations, and one easily checks that it
commutes with the gauge-invariant Hamiltonian (\ref{HTMGTdef}). The
condition $\mathcal{G}\sim0$ is a weak equality which
will be imposed as a physical state condition in the quantum field
theory, rather than as a relation among quantum operators.

To proceed with canonical quantization, we decompose form components
as $A_i=A_i^a\,T^a$ and so on, where $T^a$, $a=1,\dots,\dim G$ are the
generators of the gauge group $G$ which we take to be normalized as
$\Tr(T^a\,T^b)=\frac12\,\delta^{ab}$. The equal time canonical quantum
commutators are then given by
\beq
\left[A_i^a(x)\,,\,\Pi_j^b(y)\right]&=&\ii\delta_{ij}\,\delta^{ab}\,
\delta^{(2)}(x-y)
\label{cancomms}\eeq
with all other commutators vanishing. These imply the commutation
relations of the electric and magnetic field operators given by
\bea
\left[E_i^a(x)\,,\,E_j^b(y)\right]&=&-\frac{\ii k}{4\pi}\,
\epsilon_{ij}\,\delta^{ab}\,\delta^{(2)}(x-y) \ , \label{Ecancomms}\\
\left[E_i^a(x)\,,\,B^b(y)\right]&=&-\ii\delta^{ab}\,\epsilon_{ij}
\,\partial^j\delta^{(2)}(x-y) \ , \label{EBcancomms}\\
\left[B^a(x)\,,\,B^b(y)\right]&=&0 \ .
\label{Bcancomms}\eea

\subsection{Functional Schr\"odinger Picture\label{FSP}}

We shall work in a functional Schr\"odinger polarization wherein the
physical states are the wavefunctionals $\Psi[\mbf A]$~\cite{DJT2}. The quantum
commutators (\ref{cancomms}) are then satisfied by representing the
canonical momenta as the functional derivative operators
\beq
\Pi_i^a=-\ii\frac\delta{\delta A_i^a} \ .
\label{Piiaderivops}\eeq
Stationary states obey the functional Schr\"odinger equation $H_{\rm
  TMGT}\Psi[\mbf A]=\mathcal{E}\,\Psi[\mbf A]$ giving
\bea
&&\int_\Sigma\dd{\rm vol}_\Sigma~\left[\frac{e^2}2\,\sum_{i=1,2}\,
\sum_{a=1}^{\dim G}\left(
-\ii\frac\delta{\delta A_i^a}-\frac k{8\pi}\,\epsilon^{ij}\,
A_j^a\right)^2\right.\nn\\&&~~~~~~~~~~~~~~~~+\left.
\frac1{2e^2}\,\sum_{a=1}^{\dim G}\left(B^a
\right)^2\right]\Psi[\mbf A]~=~\mathcal{E}\,\Psi[\mbf A] \ ,
\label{fnSchreq}\eea
where $\mathcal{E}$ is the energy of the state. {\it Physical} states,
respecting the gauge symmetry, must further be annihilated by the generator
(\ref{GausslawTMGT}) of infinitesimal gauge transformations,
$\mathcal{G}\Psi[\mbf A]=0$, which gives the constraint equations
\beq
\left[\,\sum_{i=1,2}\,\sum_{b=1}^{\dim G}\left(D_{\mbf A}
\right)_i^{ab}\,\left(-\ii\frac\delta{\delta A_i^b}-
\frac k{4\pi}\,\epsilon^{ij}\,A_j^b\right)+\frac k{4\pi}\,
B^a\right]\Psi[\mbf A]=0
\label{Gausslawfn}\eeq
for each $a=1,\dots,\dim G$.

The Gauss law constraint has an immediate consequence for the structure
of the physical state wavefunctionals~\cite{AFC1}. Let us integrate
(\ref{Gausslawfn}) over the Riemann surface $\Sigma$ and substitute in
the gauge orbit parametrization (\ref{gaugeorbit}). Because of the
magnetic field term, we then find that the gauge symmetry is
represented projectively as
\beq
\Psi[\mbf A]=\e^{\ii\alpha[\bar{\mbf A},g]}~\Psi[\bar{\mbf A}] \ ,
\label{projgaugesym}\eeq
where the projective phase is a cocycle
\beq
\alpha\left[\bar{\mbf A}\,,\,g\right]=-\frac{\ii k}{4\pi}\,
\int_\Sigma\Tr\left(\bar{\mbf A}\wedge\dd g~g^{-1}\right)-
\frac k{12\pi}\,\int_{M_3}\Tr\left(g^{-1}~\dd g\right)^3
\label{projphase}\eeq
in the group cohomology of the gauge group $G$. Large gauge
transformations again change the second term in (\ref{projphase}) by the
winding number of the map $\Sigma\to G$, and so the projective phase
factor in (\ref{projgaugesym}) is well-defined only if
$k\in\zed$. This exhibits the quantization of the Chern-Simons
coefficient solely within the Hamiltonian formalism, without recourse
to any path integral description of the quantum gauge theory. The
projective cocycle is related to the chiral anomaly in two
dimensions.

Let us now solve for the physical states in the infrared limit
$e^2\to\infty$ of the topologically massive gauge theory, or more
precisely in the energy regime in which all momenta are much smaller
than the topological gluon mass (\ref{topmassTMGT})~\cite{GSST1}. The normal
ordered Hamiltonian operator (\ref{HTMGTdef}) is given in this regime
by
\beq
H_{\rm TMGT}=e^2\,\int_\Sigma\Tr\left(E^{0,1}\wedge E^{1,0}\right)
\label{HTMGTIR}\eeq
with respect to a chosen complex structure on $\Sigma$. From
(\ref{Ecancomms}) it follows that the electric
field creation and annihilation operators satisfy the commutation
relations
\beq
\left[E_z^a(z)\,,\,E_{\bz}^b(w)\right]=\frac k{2\pi}\,\delta^{ab}\,
\delta^{(2)}(z-w) \ .
\label{electriccran}\eeq
The problem of finding the physical states in the infrared limit of
topologically massive gauge theory is thereby formally a field
theoretic version of the Landau problem, exactly as we anticipated in
the previous section.

Since the Hamiltonian (\ref{HTMGTIR}) is
non-negative, the vacuum state has zero energy and is destroyed
by all of the annhilation operators,
\beq
E_z^a\Psi^{\rm vac}[\mbf A]=\left(-2\ii\frac\delta{\delta A_z^a}-\ii
A_{\bz}^a\right)\Psi^{\rm vac}[\mbf A]=0 \ .
\label{vacstatecondn}\eeq
The Gauss law (\ref{Gausslawfn}) can be written in this complex
polarization as
\beq
\left[\,\sum_{b=1}^{\dim G}\left(\left(D_{\mbf A}\right)_\bz^{ab}\,
\frac\delta{\delta A_\bz^b}+\frac k{4\pi}\,\left(D_{\mbf A}
\right)_\bz^{ab}\,A_z^b\right)-\frac k{2\pi}\,B^a\right]
\Psi^{\rm vac}[\mbf A]=0 \ .
\label{Gausslawcomplex}\eeq
This equation is related to the anomaly equation for $|k|$ flavours of
chiral fermions in two-dimensional Euclidean space~\cite{GSST1}.

The equations (\ref{vacstatecondn}) and (\ref{Gausslawcomplex}) are
simultaneously solved by a Euclidean path integral over
two-dimensional group fields as~\cite{bb,Witten4,CFKS1}
\beq
\Psi^{\rm vac}[\mbf A]=\int\DD g~\e^{-|k|\,S^+_{\rm W}[g,\mbf A]}~
\e^{-\frac{|k|}{4\pi}\,\int_\Sigma\Tr(A^{1,0}\wedge A^{0,1})} \ .
\label{vacstatesoln}\eeq
In the first factor we recognize the action (\ref{SWZNWchiral}) of the
chiral gauged WZNW model on $\Sigma$, with $\mbf A$ identified as
either $A^{1,0}$ or $A^{0,1}$ depending on the sign $k>0$ or $k<0$
of the Chern-Simons coefficient. The second factor automatically
produces the normalization factor usually required in holomorphic
quantization. It corresponds precisely to the second term in
(\ref{SCSAdef}) that was inserted by hand to ensure that the quantum gauge
theory has a well-defined classical limit. In this way we have
exhibited the appearence of induced chiral WZNW degrees of freedom on
the boundary of the membrane directly in the Hamiltonian formalism of
topologically massive gauge theory. The state $\Psi^{\rm vac}[\mbf A]$
obtained in this way is completely analogous to the Hartle-Hawking
state $\mathcal{F}[\mbf A]$ that was constructed in Section~\ref{PIF}.

\subsection{Geometrical Interpretation\label{GI}}

The construction presented in the previous subsection is closely
related to the geometric quantization of Chern-Simons gauge
theory~\cite{ADPW1}, which gives an alternative characterization of the
equivalence between the strong coupling limit of topologically massive
gauge theory and the WZNW model. Let $\mathcal{A}_\Sigma$ denote the
space of $G$-connections on the compact oriented Riemann surface
$\Sigma$ which we assume is equipped with a fixed complex
structure. Then the commutation relations (\ref{electriccran}) mean
that the electric field operators $E^{1,0}+E^{0,1}$ define a
connection of a unitary complex line bundle $\mathcal{L}^{\otimes k}$
over $\mathcal{A}_\Sigma$ of constant curvature $k\in\zed$, with
$\mathcal{L}$ the basic prequantum line bundle of geometric
quantization. The Gauss law constraint, in the form
(\ref{projgaugesym}), thereby implies that the vacuum wavefunctional
$\Psi^{\rm vac}[\mbf A]$ is a gauge-invariant {\it section} of
$\mathcal{L}^{\otimes k}$, while the ground state condition
(\ref{vacstatecondn}) implies that it is holomorphic with respect to
the canonical connection on $\mathcal{L}^{\otimes k}$.

The vector space $\mathcal{H}_\Sigma^{\rm CS}$ of holomorphic
gauge invariant sections of $\mathcal{L}^{\otimes k}$ may be presented
as the cohomology group
\beq
\mathcal{H}_\Sigma^{\rm CS}={\rm H}^0\bigl(\mathcal{P}_\Sigma^{\rm CS}\,,\,
\mathcal{L}^{\otimes k}\bigr)
\label{WSigma}\eeq
where $\mathcal{P}_\Sigma^{\rm CS}$, as in Section~\ref{PS}, is the
moduli space of flat $G$-connections on the Riemann surface
$\Sigma$. The line bundles $\mathcal{L}^{\otimes k}$ are closely
related to the Friedan-Shenker bundles of conformal field theory, and
in this way one may establish a natural isomorphism between the vector
space (\ref{WSigma}) and the corresponding finite-dimensional vector
space of WZNW conformal blocks~\cite{Witten1,Witten4}. The choice of
vacuum state (\ref{vacstatesoln}) in the space (\ref{WSigma}) thereby
characterizes a chiral WZNW correlation function on $\Sigma$.

Given this correspondence, we may now proceed to compute the amplitude
for propagation on the three-geometry $M_3=\Sigma\times[0,1]$ depicted
in Figure~\ref{Sigmax01}~\cite{CFKS1}. We insert an initial state
described by a vacuum wavefunctional $\Psi_0^{\rm vac}[\mbf A]$ of the form
(\ref{vacstatesoln}) on the lower surface $\Sigma_0$ at time
$t=0$. We then allow it to evolve in time through the bulk of the membrane,
according to the dynamics of topologically massive gauge theory, until
it reaches a final state at time $t=1$ described by a vacuum
wavefunctional $\Psi_1^{\rm vac}[\mbf A]$ on the upper surface
$\Sigma_1$ of opposite chirality to (\ref{vacstatesoln}). The
partition function of topologically massive gauge theory is thereby
determined after diagonalization as the inner product
\bea
Z_\Sigma^{\rm TMGT}&=&\bigl.\bigl\langle\Psi_0^{\rm vac}~\bigr|~
\Psi_1^{\rm vac}\bigr\rangle\nn\\&=&
\frac1{{\rm vol}\,\mathcal{G}_\Sigma}\,\int\DD\mbf A~
\overline{\DD\mbf A}~\e^{\ii S_{\rm TMGT}[A]}~
\overline{\Psi_0^{\rm vac}[\mbf A]}\,\Psi_1^{\rm vac}[\mbf A]\nn\\
&=&\sum_{\lambda,\lambda'\in(\Lambda_{G_k}^*/\Lambda^{~}_{G_k})^g}
\zeta^{\lambda\lambda'}~\psi^{~}_\lambda(\mbf m)\,
\overline{\psi}^{~}_{\lambda'}(\,\overline{\mbf m}\,) \ .
\label{statsumTMGT}\eea
Here $\Lambda_{G_k}$ denotes the root lattice of the affine Lie algebra at
level $k$ based on the gauge group $G$, so that the sums in
(\ref{statsumTMGT}) run over $g$-tuples of particular irreducible
representations of
$G$ (precisely, the integrable highest weight representations of the
current algebra at level $k$). The topological wavefunctions
$\psi^{~}_\lambda(\mbf m)$ determine holomorphic conformal blocks and
they span the Hilbert space $\mathcal{H}_\Sigma^{\rm
  CS}$~\cite{BN1,LR1}. They depend on the complex moduli $\mbf
m\in\complex^{3g-3}$ of the Riemann surface $\Sigma$ and can be
computed from the lowest Landau level wavefunctions for the Landau
problem on $\Sigma$~\cite{Kogan5}, as described in
Section~\ref{HS}. They are succinctly expressed in this way in terms
of holomorphic genus $g$ Jacobi theta-functions at level
$k$~\cite{CFKS1}.

The coupling coefficients $\zeta^{\lambda\lambda'}\in\nat_0$ ensure that
the bilinear form (\ref{statsumTMGT}) is a modular invariant of the
string worldsheet $\Sigma$~\cite{CIZ1}. In most cases the partition
function will be given by a diagonal sum,
$\zeta^{\lambda\lambda'}=\delta^{\lambda\lambda'}$, but there can be
examples of three-manifolds $M_3$ for which the inner product assumes
an entangled form. The complete three-dimensional version of the
standard ADE classification of rational conformal field
theories~\cite{CIZ2} is not known, and it presumably involves the
gravitational sector of the topological membrane, to be described in
the next section. As is evident from the second line of
(\ref{statsumTMGT}), the modular invariant statistical sum for the
topological membrane contains boundary degrees of freedom, appearing
as in (\ref{vacstatesoln}), which ensure bulk gauge invariance of the
inner product and also that classical extrema contribute to the path
integral governing the quantum theory of the membrane~\cite{CFKS1}.

However, in spite of what is written here, there is not quite a complete
holomorphic factorization into left-moving and right-moving worldsheet
degrees of freedom. There are subtleties in arriving at such a
factorization for generic gauged WZNW models which are avoided by constructing
vacuum wavefunctionals involving {\it two} gauge connections on
$\Sigma$~\cite{Witten4}. We will not enter any further into this
discussion here.

\section{Topologically Massive Gravity\label{TMG}}
\setcounter{equation}{0}

Because the Yang-Mills term in the topologically massive gauge theory
action (\ref{STMGTAdef}) couples to the spacetime metric of $M_3$,
radiative corrections will generate three-dimensional gravitational
terms. The proper formulation of the membrane quantum theory must
thereby include a sum over all metrics weighted by the appropriate
gravity actions. In this section we will briefly describe the
gravitational sector of the topological membrane theory and the
ensuing emergence of the string dilaton field. For the most part, in
subsequent sections we will ignore gravitational contributions, but
here we include a quick description for completeness.

\subsection{Conformal Coupling and the Dilaton\label{CCD}}

To incorporate gravitational terms into the action (\ref{STMGTAdef}),
we will work in the first order formalism of general relativity. The
fibers of the frame bundle over the three-manifold $M_3$ are spanned by
local triad fields $e^a\in\Omega^1(M_3)$, $a=1,2,3$ which together
transform as a vector under local $SO(2,1)$ Lorentz
transformations. The frame bundle carries a
canonical spin-connection $\omega$ transforming as a gauge connection
under the local $SO(2,1)$ group, which is torsion-free, compatible with the
metric of $M_3$, and has curvature
\beq
R^a(\omega)=\dd\omega^a+\epsilon^{abc}\,\omega^b\wedge\omega^c
\label{Raomegadef}\eeq
with $\epsilon^{123}=+\,1$. The compatibility condition on $\omega$
means that the triads $e^a$ are not independent variables. Let us
further introduce a dimensionless scalar field $D$ on $M_3$.

The action for the conformal coupling of topologically massive gauge
theory to topologically massive gravity is defined by the non-local
functional~\cite{KMS1}
\bea
S_{\rm CTMGT}[A,\omega;D]&=&\int_{M_3}\biggl[\kappa\,D^2\,e^a\wedge
R^a(\omega)+8\kappa~\dd D\wedge*\,\dd D\biggr.\nn\\&&~~~~~~~~~~-
\left.\frac1{e^2D^2}\,
\Tr\left(F_A\wedge*F_A\right)\right]\nn\\&&-\,8\kappa\,\int_{\partial M_3}
D*_2\,i^{~}_{\partial_\perp}\,\dd D+S_{\rm CS}[A,\omega] \ ,
\label{SCTMGTdef}\eeq
where
\bea
S_{\rm CS}[A,\omega]&=&\int_{M_3}\left[\frac k{4\pi}\,\Tr\left(
A\wedge\dd A+\mbox{$\frac23$}\,A\wedge A\wedge A\right)\right.\nn\\&&
+\left.\frac{k'}{8\pi}\left(\omega^a\wedge\dd\omega^a+
\mbox{$\frac23$}\,\epsilon^{abc}\,\omega^a\wedge\omega^b\wedge\omega^c
\right)\right]
\label{SCSgaugegrav}\eeq
is the sum of the gauge and gravitational Chern-Simons
actions~\cite{DJT1,Witten1}. In
(\ref{SCTMGTdef}), $\kappa$ is the three-dimensional Planck mass and
the first term is a modification of the usual Einstein-Hilbert action.
$i^{~}_{\partial_\perp}:\Omega^p(M_3)\to\Omega^{p-1}(M_3)$ denotes
interior multiplication by the
vector field $\partial_\perp$ which locally spans the one-dimensional
fibers of the normal bundle $N\Sigma$ to the boundary $\Sigma=\partial
M_3$ in $M_3$. Since the non-compact Lie group $SO(2,1)$ is
contractible, there is no quantization condition that needs to be
imposed on the gravitational Chern-Simons coefficient in
(\ref{SCSgaugegrav}) and one may choose any $k'\in\real$.

The boundary term in (\ref{SCTMGTdef}) ensures that the action is
invariant under three-dimensional conformal transformations when $M_3$
has a non-empty boundary~\cite{KMS1}. We have suppressed for simplicity the
boundary terms required to give the corresponding quantum field theory
a well-defined classical limit. In the phase with a non-vanishing vacuum
expectation value $\langle D^2\rangle\neq0$, corresponding to a vacuum
with spontaneously broken conformal symmetry, one can gauge the scalar
field $D$ away by a Weyl transformation of the metric with conformal
factor $\Omega$ given by $\Omega^2=\langle D^2\rangle/D^2$. Then the
resulting action (\ref{SCTMGTdef},\ref{SCSgaugegrav}) describes
topologically massive gravity coupled minimally to topologically
massive gauge theory. This phase contains a propagating graviton mode
of topological mass
\beq
\mu'=\frac{8\pi\,\kappa\,\left\langle D^2\right\rangle}{|k'|}
\label{topgravmassD}\eeq
along with a propagating gluon of topological mass
\beq
\mu=\frac{e^2|k|\,\left\langle D^2\right\rangle}{4\pi} \ .
\label{topgluonmassD}\eeq
The quantum fluctuations of the scalar field $D$ thereby set the mass
scales of the bulk theory.

Let us now turn off the gauge sector of the theory and set
$A=0$. Following the analogous procedures to those used before, the
bulk quantum field theory defined by
(\ref{SCTMGTdef},\ref{SCSgaugegrav}) can be shown to induce a
two-dimensional gravity action on the boundary $\Sigma=\partial
M_3$~\cite{Carlip1}--\cite{KMS1},\cite{Ashworth1}. We refer to this
two-dimensional quantum gravity as a ``deformed'' Liouville theory,
and it is described by the action functional~\cite{KMS1}
\bea
S_{\rm L}[D,\phi]&=&\int_\Sigma\left[-\frac1{4\pi}\,\left(\ln
D^4+\phi\right)\,R^{(2)}-2\kappa\,D*_2\,i^{~}_{\partial_\perp}\,\dd D
\right.\nn\\&&+\left.\frac1{16\pi}\,\dd\phi\wedge*_2\,\dd\phi+\Lambda
*_2\e^{-\phi}\right] \ ,
\label{defLiouville}\eea
where $\phi$ is the Liouville field and $R^{(2)}$ the curvature
two-form of $\Sigma$. The cosmological constant is
determined by the topological graviton mass (\ref{topgravmassD})
as~\cite{Kogan2}
\beq
\Lambda=\left(\mu'\,\right)^2 \ .
\label{cosmconst}\eeq
The appearence of a Liouville field theory from topologically massive
gravity is not entirely surprising, given that the boundary theory can be
formulated in terms of an $SL(2,\real)$ WZNW model~\cite{Alek1}, while
the bulk theory is formulated in terms of an $SO(2,1)$ Chern-Simons
gauge theory. The natural identifications of the two models can now be
heuristically deduced from the group isomorphism
\beq
SO(2,1)=PSL(2,\complex)=SL(2,\real)\times SL(2,\real) \ ,
\label{SO21iso}\eeq
inducing both chiralities of the $SL(2,\real)$ worldsheet theory as in
(\ref{defLiouville}). The gravitational dressing of conformal field
theories by topologically massive gravity is studied
in~\cite{A-CKS1,KogSz1}.

In addition to inducing the gravitational sector of the string theory,
we see from (\ref{defLiouville}) that the scalar field $D$ can be
naturally identified as the three-dimensional version of the string
dilaton field. In particular, the string coupling is given by
\beq
g_s=\left\langle D^4\right\rangle \ .
\label{gsD4}\eeq
Let us summarize a few of the generic features of the identification
of the dilaton in this way~\cite{KMS1}:
\begin{enumerate}
\item In string theory, the target space tachyon operator usually
  depends on the dilaton, which can be used to change $g_s$ and thus
  it controls the scale transformation properties of the strings. This
  is consistent with the manner in which the field $D$ sets the bulk
  mass scales in (\ref{topgravmassD}) and (\ref{topgluonmassD}), and
  this feature is important for correctly formulating T-duality within
  the framework of topological membranes.
\item The dilaton in the topological membrane approach has a nice
  dynamical origin in terms of the fluctuating geometry and
  propagating bosons in the bulk three-manifold.
\item With suitable boundary conditions for the field $D$ on $\partial
  M_3$, one can also generate a {\it dynamical} string coupling
  $g_s$ through the identification (\ref{gsD4}). This is reminescent
  of the relationship between M-Theory and Type~IIA superstring
  theory, and suggests that it could provide a possible worldsheet
  origin for the eleventh spacetime dimension of M-Theory.
\end{enumerate}

\subsection{Hamiltonian Quantization\label{HQTMG}}

Let us now analyse the Schr\"odinger wavefunctionals for pure
topologically massive gravity ($A=D=0$ in
(\ref{SCTMGTdef},\ref{SCSgaugegrav})) on the three-manifold
$M_3=\Sigma\times[0,1]$. We begin with the infrared limit
of the theory, $k'\to0$, in which the massive graviton decouples from
the spectrum and the theory reduces to pure Einstein gravity in
three-dimensions which is a topological field
theory~\cite{Witten5}. Let us consider the string worldsheet which is
a torus $\Sigma={\mathbb T}^2$ of
modulus $\tau=\tau_1+\ii\tau_2$, $\tau_2>0$. With $J_a$, $a=1,2,3$
denoting the generators of $SO(2,1)$, the topological wavefunctions then depend
on the mean extrinsic curvature $K$ of $\Sigma$ in $M_3$ and two commuting
$SO(2,1)$ holonomies $\e^{\lambda_1\,J_2}$, $\e^{\lambda_2\,J_2}$
as~\cite{Carlip2}
\beq
\Psi_{\rm grav}(\lambda_1,\lambda_2;K)=\int_{\mathcal{F}}
\frac{\dd^2\tau}{\tau_2^2}~\frac{\lambda_1-\tau\,\lambda_2}{\pi\,
\tau_2^{1/2}\,K}~\e^{-\ii|\lambda_1-\tau\,\lambda_2|^2/\tau_2K}
{}~\chi(\tau,\overline{\tau}\,) \ ,
\label{Psigravgen}\eeq
where $\mathcal{F}$ is a fundamental modular domain of the upper
complex half-plane, and the Schr\"odinger equation is equivalent to
the requirement that the function $\chi(\tau,\overline{\tau}\,)$ be
an automorphic Maass form of modular weight $\frac12$.

The ground state corresponds to the choice
\beq
\chi^{(0)}(\tau,\overline{\tau}\,)=\tau_2^{1/2}\,\eta(\tau)^2
\label{chi0choice}\eeq
in (\ref{Psigravgen}), where
\beq
\eta(\tau)=\e^{\pi\ii\tau/12}\,\prod_{n=1}^\infty
\left(1-\e^{2\pi\ii n\,\tau}\right)
\label{Dedekind}\eeq
is the Dedekind function. We can now work out the gravitational analog
of the membrane inner product which we defined in Section~\ref{GI},
and one finds
\beq
\left.\left\langle\Psi_{\rm grav}^{(0)}~\right|~\Psi_{\rm grav}^{(0)}
\right\rangle=\int_{\mathcal
  F}\frac{\dd^2\tau}{\tau_2^2}~\bigl|\eta(\tau)\bigr|^4 \ .
\label{ghostcontr}\eeq
This result coincides exactly with the diffeomorphism ghost
contribution to the string theory torus partition
function.

Reinstating the gauge sector of the bulk theory, the full
transition amplitude for the propagation of states between the
boundaries $\Sigma_0$ and $\Sigma_1$ is then given using the result of
Section~\ref{GI} as
\beq
Z_{\rm grav}=\int_{\mathcal
  F}\frac{\dd^2\tau}{\tau_2^2}~\bigl|\eta(\tau)\bigr|^4\,
\sum_{\lambda,\lambda'\in(\Lambda_{G_k}^*/\Lambda^{~}_{G_k})^g}
\zeta^{\lambda\lambda'}~\overline{\psi}^{~}_{\lambda}(\,\overline{\tau}\,)\,
\psi^{~}_{\lambda'}(\tau) \ .
\label{Zgrav}\eeq
This amplitude contains the correct integration over moduli space,
which is induced by the integration over the holonomy parameters
$\lambda_1$ and $\lambda_2$ in the overlap (\ref{ghostcontr}) of
initial and final gravitational wavefunctions. Note that
the two dimensional ghosts do {\it not} emerge from gauge fixing
the gravitational diffeomorphisms in three-dimensions, as these
bulk ghost fields completely decouple on the three-manifold
$M_3=\Sigma\times[0,1]$~\cite{CK3}. The proper description of string
ghost fields in three-dimensional language is not presently known. A
related problem is the expression of the worldsheet BRST formalism in
the topological membrane. The BRST conditions are presumably related
to the loop equations of the bulk gauge theory.

This is the only detailed result concerning the gravitational
wavefunctionals that is presently available. However, from the
structure of the gravitational action written in a suitable
parametrization, it is possible to show, in a manner analogous to what
we did in Section~\ref{HS}, that the physical Hilbert space of the
full topologically massive gravity theory
($\mu'=8\pi\,\kappa/|k'|<\infty$) on $\Sigma\times[0,1]$ is the
product~\cite{CK2,Kogan6}
\beq
\mathcal{H}_\Sigma^{\rm TMG}=\mathcal{H}_{\rm f}[\varphi]\otimes
\left(\mathcal{H}_{\mbf m}\right)^{\otimes(3g-3)} \ .
\label{HTMG}\eeq
The first factor is the Hilbert space of the propagating
graviton of mass $\mu'$, while the second factors are topological
contributions from quantum mechanical degrees of freedom induced by
the moduli $\mbf m\in\complex^{3g-3}$ of the Riemann surface
$\Sigma$.

\section{Wilson Lines\label{WL}}
\setcounter{equation}{0}

In this section we will further deform the topological Chern-Simons
gauge theory by coupling it to charged matter fields in the bulk. The
consequences of this will be many-fold. Adding charged matter
corresponds to deforming the corresponding conformal field theory and
allows us to describe the simplest vertex operators that create
primary string states. Later on, the careful incorporation of such a
deformation will be used to describe the boundary states that
correspond to D-branes.

\subsection{Deformations of Conformal Field Theories\label{DCFT}}

Suppose that we {\it deform} our worldsheet conformal field theory,
described generically by an action $S_{\rm CFT}$, by some collection
of operators $V_I$, $I\in\mathcal{I}$ to produce an
action~\cite{Zam1,LC1}
\beq
S_\Lambda=S_{\rm CFT}+\sum_{I\in\mathcal{I}}\lambda^I(\Lambda)\,\int_\Sigma
\dd^2z~V_I(z,\bz\,) \ ,
\label{Sdeform}\eeq
where the couplings $\lambda^I(\Lambda)$ generally depend on some worldsheet
scale $\Lambda$. For a relevant deformation, the original conformal
field theory with action $S_{\rm CFT}$ is described as an ultraviolet
fixed point of the corresponding renormalization group flows, while
for an irrelevant deformation it corresponds to an infrared fixed
point. It is possible to have renormalization group flows which
connect two different conformal field theories.

In the topological membrane approach, let us now consider the addition
of charged matter in the bulk, so that even the pure Chern-Simons
theory is no longer topological, as there are local propagating degrees of
freedom. This will induce a deformed two-dimensional conformal field
theory on the boundary. To see this, consider for simplicity the case of a
single-charge deformation of the worldsheet theory, whose fields we
denote collectively by $\Phi$. The (suitably normalized) partition
function is given by
\bea
Z_{\rm ws}&=&\int\DD\Phi~\e^{-S_{\rm CFT}+\lambda\,\int_\Sigma
\dd^2z~V(z,\bz\,)}
\nn\\&=&1+\sum_{n=1}^\infty\frac{\lambda^n}{n!}\,\int_\Sigma\dd^2z_1~
\cdots\int_\Sigma\dd^2z_n~\bigl\langle V(z_1,\bz_1)\cdots
V(z_n,\bz_n)\bigr\rangle^{~}
\label{Zwsdef}\eeq
where the averages denote $n$-point correlation functions in the
unperturbed conformal field theory. In WZNW models, correlators of
primary vertex operators generically exhibit a holomorphic
factorization into chiral and anti-chiral conformal blocks as
\beq
\left\langle\,\prod_{i=1}^nV_{\lambda_i}(z_i,\bz_i)\right\rangle=
\left\langle\,\prod_{i=1}^nV^{\rm L}_{\lambda_i}(z_i)\right\rangle\,
\left\langle\,\prod_{i=1}^nV^{\rm R}_{\lambda_i}(\,\bz_i)\right\rangle
\label{corrsholfact}\eeq
where $\lambda_i$ label irreducible representations of the group
$G$ of the WZNW model. More precisely, the left-hand side of
(\ref{corrsholfact}) is actually a sum over left-right conformal
blocks, but to avoid clutter we simply write the factorization as
displayed.

Let us now consider the membrane geometry $M_3=\Sigma\times[0,\beta]$,
where we give the time direction an arbitrary length
$\beta\in\real$. Then the conformal field theory correlation functions
are induced by {\it open} (gauge non-invariant) Wilson lines
\beq
W^{\{\lambda_i\}}[A]=W_{\bigcup_iC_i(\lambda_i)}[A]=
\prod_{i=1}^n\Tr^{~}_{\lambda_i}\,{\rm P}\,
\exp\left(\ii\int_{C_i}A\right) \ ,
\label{openWilsonlines}\eeq
where $C_i$ for each $i=1,\dots,n$ is an oriented, vertical open
contour in $M_3$ with endpoints $z_i\in\Sigma_0$ and $\bz_i\in\Sigma_1$ (see
Figure~\ref{Sigmax01}). The insertion of the operator
(\ref{openWilsonlines}) in the bulk gives the Aharonov-Bohm phase
factor (holonomy) for propagating charged particles, minimally coupled to the
gauge field $A$, which move from left to right worldsheets. In other
words, a gas of (open) Wilson lines describes charged matter in the
bulk. More precisely, the operators (\ref{openWilsonlines}) are really
Polyakov lines at inverse finite temperature~$\beta$, the internal
size of the topological membrane. We have thereby derived a remarkable
new aspect of our underlying holographic correspondence, that charged
three-dimensional matter at finite temperature is equivalent to a
deformed two-dimensional conformal field theory. The bulk parameter
$\beta$ plays the role of the fugacity $\lambda$ of (\ref{Zwsdef}) in
three dimensions~\cite{Kogan3}.

\subsection{Polyakov Loops in Topologically Massive Gauge
  Theory\label{PLTMGT}}

Given the correspondence we arrived at in the previous subsection, let
us now study in more detail the Polyakov loop operators of
topologically massive gauge theory. We will use them to study {\it
  chiral} aspects of the induced boundary conformal field theory~\cite{bb}. For
this, we choose a gluing automorphism which identifies the left and
right worldsheets $\Sigma_0=\Sigma_1$, and thereby study gauge
dynamics on the membrane geometry $M_3=\Sigma\times\sphere^1$. The
Polyakov loop operators (\ref{openWilsonlines}) then correspond to
characters of the pure gauge parts $g$ of the gauge field $A$
(see~(\ref{gaugeorbit})) in representations $\lambda_i$ defined by
\beq
\chi^{~}_{\lambda_i}(z_i)=\Tr^{~}_{\lambda_i}\bigl(g(z_i)\bigr)=
\Tr\bigl[g(z_i)\bigr]_{\lambda_i} \ .
\label{chilambdaizi}\eeq
We will consider again the infrared limit $e^2\to\infty$ whereby the
vacuum amplitude of the gauge theory gives the pure Chern-Simons
theory partition function at finite temperature, which may be computed
as the thermal average
\beq
Z_\Sigma^{\rm CS}=\Tr^{~}_{\mathcal{H}_\Sigma^{\rm CS}}\left(
\e^{-\beta\,H_{\rm CS}}\right) \ .
\label{thermalpartfn}\eeq
Because Chern-Simons theory is a topological field theory, it has a
vanishing Hamiltonian, $H_{\rm CS}=0$, and the partition function
(\ref{thermalpartfn}) thereby computes the dimension of the
corresponding Hilbert space $\mathcal{H}_\Sigma^{\rm CS}$,
\beq
Z_\Sigma^{\rm CS}=\dim\mathcal{H}_\Sigma^{\rm CS} \ ,
\label{dimCSHilbert}\eeq
independently of the internal membrane size $\beta$~\cite{BT1}.

The Chern-Simons Hilbert spaces over punctured Riemann surfaces
correspond to non-dynamical external charged particles in the bulk, or
equivalently to Polyakov loop insertions in the path
integral~\cite{Witten1}. The associated vacuum wavefunctionals
generalizing (\ref{vacstatesoln}) are given by~\cite{bb,GSST1}
\bea
\Xi^{\{\lambda_i\}}\bigl[\{z_i\}\,;\,\mbf A\bigr]&=&
\e^{-\frac{|k|}{4\pi}\,\int_\Sigma\Tr(A^{1,0}\wedge A^{0,1})}\nn\\&&
\times\,\int\DD g~\bigotimes_{i=1}^n\bigl[g(z_i)\bigr]_{\lambda_i}~
\e^{-|k|\,S_{\rm W}^+[g,\mbf A]} \ .
\label{Xilambdaidef}\eea
The representations $\lambda_i$ act on complex vector spaces $V_i$,
$i=1,\dots,n$. Thus the wavefunctional (\ref{Xilambdaidef}) may be
regarded as an operator on the product of the corresponding
representation spaces $V_i$, and one has
\beq
\Xi^{\{\lambda_i\}}\bigl[\{z_i\}\,;\,\mbf A\bigr]~\in~\mbf V=
\bigotimes_{i=1}^n\left(V_i^*\otimes V_i^{~}\right) \ .
\label{Xirepspaces}\eeq

Under a local gauge transformation (\ref{gaugeorbit}), these states
transform under a projective representation of the gauge group as
\beq
\Xi^{\{\lambda_i\}}\bigl[\{z_i\}\,;\,\mbf A\bigr]=\e^{\ii
\alpha[\bar{\mbf A},g]}~\Xi^{\{\lambda_i\}}\bigl[\{z_i\}\,;\,
\bar{\mbf A}\bigr]\otimes\bigotimes_{i=1}^n\bigl[g(z_i)
\bigr]_{\lambda_i} \ ,
\label{Xigaugetransf}\eeq
where the projective phase $\alpha[\bar{\mbf A},g]$ is again the group
cocycle given by (\ref{projphase}). Thus the wavefunctionals
(\ref{Xilambdaidef}) behave analogously to gauge theory correlators of
the {\it open} Wilson lines
(\ref{openWilsonlines}). Furthermore, the property of being
annihilated by the electric field operators $E^{1,0}$ of
Section~\ref{HQ} is clearly insensitive to insertions of the
group-valued field $g$ on the string worldsheet~$\Sigma$. Thus the
wavefunctionals (\ref{Xilambdaidef}) still define vacuum states of the
infrared limit of topologically massive gauge theory for the membrane
geometry $M_3=\Sigma\times\sphere^1$.

\subsection{The Verlinde Formula\label{VF}}

The computation of the dimensions (\ref{dimCSHilbert}) proceeds by
introducing an inner product, as in Section~\ref{GI}, appropriate to the states
(\ref{Xilambdaidef}). Under the holographic correspondence, this
produces conformal field theory correlators of the corresponding
character insertions (\ref{chilambdaizi}) on $\Sigma$, and a careful
treatment of the resulting gauge theory path integral~\cite{bb,BT1}
reproduces exactly the anticipated Verlinde formula for the dimensions
of the spaces of conformal blocks on an $n$-punctured Riemann surface
$\Sigma$ of genus $g$~\cite{Verlinde1}
\bea
\left\langle\,\prod_{i=1}^n\chi^{~}_{\lambda_i}(z_i)\right\rangle&=&
\frac1{{\rm vol}\,\mathcal{G}_\Sigma}\,\Tr^{~}_{\mbf V}
\left.\left\langle\Xi^{\{\lambda_i\}}\{z_i\}~\right|~
\Xi^{\{\lambda_i\}}\{z_i\}\right\rangle\nn\\&=&\sum_{\lambda\in
\Lambda_{G_k}^*/\Lambda_{G_k}^{~}}\left({\sf S}_{0\lambda}\right)^{2-2g-n}
\,\prod_{i=1}^n{\sf S}_{\lambda_i\lambda} \ .
\label{Verlindeformula}\eea
Here ${\sf S}_{\lambda\lambda'}$ is the modular $\sf S$-matrix of the
gauge group $G$, which we assume is compact, connected and simple,
given by the formula
\beq
{\sf S}_{\lambda\lambda'}=\sqrt{\frac{(-1)^{|\Delta_+|}}{(k+c_v)^r}\,
\frac{{\rm vol}\,\Lambda^*_{G_k}}{{\rm vol}\,\Lambda^{~}_{G_k}}}~
\sum_{w\in\mathcal{W}_{G_k}}{\rm sgn}(w)~\e^{-\frac{2\pi\ii}{k+c_v}\,
(\lambda+\rho\,,\,w(\lambda'+\rho))}
\label{modularSmatrix}\eeq
with $|\Delta_+|$ the number of positive roots, $r$ the rank,
$\mathcal{W}_{G_k}$ the Weyl group, $c_v$ the quadratic Casimir in the
adjoint representation, and $\rho=\frac12\,\sum_{\alpha\in\Delta_+}\alpha$ the
Weyl vector of the current algebra based on $G$ at level $k$. The
quadratic form in (\ref{modularSmatrix}) is
inherited from the inner product on the root lattice. When
$\Sigma={\mathbb T}^2$ ($g=1$, $n=0$), the formula
(\ref{Verlindeformula}) also computes the number of integrable
highest weight representations of the affine Lie algebra associated to
$G$ at level $k$. The independence of this correlation function on the puncture
positions $z_i\in\Sigma$ owes to the topological invariance of the
quantum gauge theory in the infrared limit.

\begin{figure}[ht]
\centerline{\epsfxsize=5cm
\epsfbox{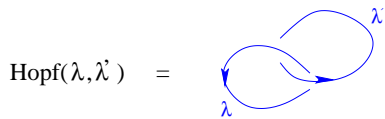}}
\caption{The Hopf linking of two charges $\lambda$ and $\lambda'$.}
\label{HopfLink}\end{figure}

The $\sf S$-matrix has a very explicit description in
three-dimensional terms via the formalism of surgery constructions on
three-manifolds~\cite{Witten1}. It can be represented in terms of the
vacuum expectation value
of the Wilson line operator (\ref{openWilsonlines}) taken around the
components of the Hopf link (Figure~\ref{HopfLink}) carrying charges
$\lambda,\lambda'$, with respect to the Chern-Simons path integral
defined on the three-sphere $\sphere^3$, as
\beq
{\sf S}_{\lambda\lambda'}=\bigl\langle W_{{\rm
    Hopf}(\lambda,\lambda')}^{~}\bigr\rangle^{~}_{\sphere^3} \ .
\label{SmatrixHopf}\eeq
This result is derived by exploiting the fact that the path integral
defines a modular functor to cut out a tubular neighbourhood
surrounding the link, giving the partition function of Chern-Simons
gauge theory on a solid torus, and then gluing it back after
performing a modular transformation along its boundary
${\mathbb T}^2$. With it one arrives at a purely geometrical and dynamical
derivation of the Verlinde diagonalization formula. However, the proper way to
encode the fusion rules of the underlying rational conformal field
theory, upon which the Verlinde formula is based~\cite{Verlinde1}, within the
Chern-Simons framework is as yet an unsolved problem in the general
case. This problem has been addressed from the point of view of the
Chern-Simons inner product in~\cite{FG1}--\cite{GT-N-B1}. Some
progress has been made in the axiomatic approaches to
three-dimensional topological quantum field theories~\cite{FFFS1}.

It is instructive to consider the simplest instance of gauge group
$G=U(1)$, in which case the description of the fusion rules in
three-dimensional terms is completely understood~\cite{bb}. When the
Chern-Simons coefficient is
a rational number $k=4p/q$, with $p$ and $q$ relatively prime positive
integers, the holographic dual of the three-dimensional theory is the
$c=1$ conformal field theory of the rational circle, with its extended
chiral algebra~\cite{BN1,CFKS1,bb}. The modular $\sf S$-matrix in this
case is given by
\beq
{\sf
  S}_{\lambda\lambda'}=\frac1{\sqrt{pq}}~\e^{-2\pi\ii\lambda\,\lambda'/pq} \ ,
\label{SmatrixU1}\eeq
where $\lambda,\lambda'\in\zed_{pq}$ generate a Narain lattice of
charges~\cite{Narain1,NSW1}. These matrix elements can be thought of
in three-dimensional
terms entirely by computing the correlation function on the
three-sphere for a generic Wilson loop (\ref{openWilsonlines})
consisting of $n$ connected components $C_i$ corresponding to the
periodic worldlines of particles with charges $\lambda_i$. In this
simple case the result can be expressed very explicitly
as~\cite{Polyakov1}
\beq
\bigl\langle W^{\{\lambda_i\}}\bigr\rangle^{~}_{\sphere^3}=
\prod_{i,j=1}^n\e^{-\frac{2\pi\ii}k\,\lambda_i\lambda_j\,
\#(C_i,C_j)} \ ,
\label{WilsoncorrU1}\eeq
where $\#(C_i,C_j)$ is the linking number which counts the number of
signed intersections of $C_i$ with the surface spanned by $C_j$. The
phase factors in (\ref{WilsoncorrU1}) can be interpreted as
Aharonov-Bohm phases arising from the circulation of one
charged particle about another. In particular, this formula expresses
the equivalence between the anomalous spins
\beq
\Delta=\frac{\lambda^2}k
\label{anomspins}\eeq
in the dual two-dimensional and three-dimensional
theories~\cite{A-CKS1}.

\section{Compactification\label{Compact}}
\setcounter{equation}{0}

In this section we will briefly describe how the Narain lattice
containing the allowed spectrum of string charges arises in the
topological membrane framework. The construction presented in the
following quickly generalizes to any toroidal compactification of the
target space, but for simplicity we will only describe the case of a
spacetime compactified on a circle $\sphere^1$. The proper
description in the case of strings propagating in generic group
manifolds is not presently understood.

\subsection{Monopole-Instanton Operators\label{M-IO}}

To incorporate the Narain lattice of charges, we need to construct
string winding modes, which in turn requires some process that leads
to charge non-conservation in topological membrane theory. Such
processes arise due to the presence of monopole-instantons in {\it
  compact} $U(1)$ Chern-Simons gauge
theory~\cite{Luscher1}--\cite{KK1}, which are remnants of
't~Hooft-Polyakov monopoles in a spontaneously broken $SU(2)$
Chern-Simons-Higgs model and lead to the possibility of tunneling
between states of different magnetic flux or charge. These topological
defects also arise in compact $U(1)$ topologically massive gauge
theory and lead to a BKT phase transition on the string
worldsheet~\cite{Sath1}--\cite{AK1}.

In $U(1)$ topologically massive gauge theory on the three-geometry
$M_3=\Sigma\times[0,1]$, the Chern-Simons coefficient $k$ is related
to the compactification radius $R$ of the target space through
\beq
|k|=\frac{4R^2}{\alpha'} \ ,
\label{kRrel}\eeq
where $\alpha'$ is the string slope. Restricting to a {\it compact}
$U(1)$ gauge group means that the pure gauge part $g$ of the gauge
connection $A$ in (\ref{gaugeorbit}) is a compact, $\sphere^1$-valued
field on the string worldsheet $\Sigma$. Thus the compactification of
the gauge group yields the desired target space compactification, and
also the appropriate topological field configurations that generate
the Narain lattice, as we now proceed to demonstrate.

{}From the Gauss law (\ref{GausslawTMGT}) it follows that the elements
of the gauge group are the operators
\beq
U=\exp\left[-\ii\int_\Sigma\theta*_2\left
(*_2\,\dd\,*_2 E-\frac k{4\pi}\,B-\rho\right)\right] \ ,
\label{gaugegpelts}\eeq
where $\rho$ is the charge density of external minimally coupled bulk
matter and the quantity in the second set of brackets is the gauge
group generator $\mathcal G$. They generate the infinitesimal
time-independent gauge transformations
\beq
U\,\mbf A\,U^{-1}=\mbf A+\dd\theta \ ,
\label{Ugaugetransf}\eeq
so that the physical states $|\Psi\rangle$
of the gauge theory are those which are invariant under their action
on the quantum Hilbert space,
$U|\Psi\rangle=|\Psi\rangle$. In the presence of topologically
non-trivial gauge field configurations we must also take into account
large gauge transformations. They can be incorporated by taking the
gauge function $\theta(z)$ in (\ref{gaugegpelts}) to be the
multi-valued angle function of the Riemann surface $\Sigma$~\cite{BS1}
\beq
\theta_{z_0}(z)={\rm Im}\,\ln\left(\frac{\mbf E_{z_0}(z)}{\mbf E_{z'}(z)\,
\mbf E_{z_0}(z')}\right) \ ,
\label{anglefndef}\eeq
giving the angle between $z$ and some reference point $z_0$ on
$\Sigma$. Here $z'$ is an arbitrary but fixed reference point on
$\Sigma$, while $\mbf E_{z_0}(z)$ denotes the prime form of $\Sigma$.

Via an integration by parts in (\ref{gaugegpelts}) we thereby discover
the extra local operators~\cite{KK1,Marino1,CKL1}
\bea
V(z_0)&=&\exp\left[-\ii\int_\Sigma\left(E+\frac k{4\pi}\,*_2A\right)
\wedge*_2\,\dd\ln\mbf
E_{z_0}\right.\nn\\&&+\left.\ii\int_\Sigma\theta_{z_0}*_2\rho\right] \
{}.
\label{extralocalops}\eea
These operators commute with non-compact gauge transformations, and
the physical state conditions $V(z_0)|\Psi\rangle=|\Psi\rangle$ can be
imposed simultaneously for all $z_0\in\Sigma$. From the
canonical commutation relations (\ref{Ecancomms})--(\ref{Bcancomms})
it follows that
\beq
\bigl[B(z)\,,\,V^n(z_0)\bigr]=2\pi\,n\,\delta^{(2)}(z-z_0)~
V^n(z_0) \ ,
\label{BVncancomm}\eeq
and so the operator $V^n(z_0)$ creates a point-like magnetic vortex at
$z_0\in\Sigma$ of flux
\beq
\frac1{2\pi}\,\int_\Sigma F_{\mbf A}=n \ ,
\label{vortexflux}\eeq
which is the monopole number of the topologically non-trivial gauge
field configuration $\mbf A$.

Furthermore, we can integrate the Gauss law (\ref{GausslawTMGT}) and
use the exponential decay of the electric field $E$ at large distance
scales, owing to the topological mass of the photon in the theory. It
then follows that the operator $V^n(z_0)$ also carries an electric
charge
\beq
\Delta Q=-\frac{n\,k}2 \ .
\label{DeltaQVn}\eeq
This charge is unobservable in the long wavelength limit far from the
vortex because of the exponential fall-offs, and the Aharonov-Bohm
linking phases are all equal to $1$~\cite{CKL1}. We conclude that
$V^n(z_0)$ is a monopole-instanton operator, creating a dyon that
interpolates between topologically inequivalent vacua of the
topologically massive gauge theory.

\subsection{The Narain Lattice\label{SWM}}

The monopole-instantons also shift the allowed charge spectrum of the
quantum field theory. From (\ref{projgaugesym}) it follows that, in
the functional Schr\"odinger picture, the physical states acquire a
projective phase under gauge transformations as
\beq
V(z_0)\Psi[\mbf A]=\e^{\ii\int_\Sigma\theta_{z_0}*_2(\frac k{8\pi}\,B-
\rho)}~\Psi[\mbf A+\dd\theta_{z_0}] \ .
\label{Vz0projphase}\eeq
Now single-valuedness of the gauge cocycle in (\ref{Vz0projphase})
under periodic shifts $\theta_{z_0}\to\theta_{z_0}+2\pi$ of the angle
function requires the quantization condition
\beq
Q=\int_\Sigma*_2\rho=m+\frac k{8\pi}\,\int_\Sigma F_{\mbf A} \ ,
\label{chargequant}\eeq
where the integer $m$ can be interpreted as a particle winding number
around the monopole-instanton at $z_0\in\Sigma$. The condition
(\ref{chargequant}) generalizes the usual Dirac charge quantization of
compact quantum electrodynamics (recovered in the limit $k=0$). From
the flux quantization condition (\ref{vortexflux}) it follows that the
spectrum of electric charges is thereby given as
\beq
Q=m+\frac{n\,k}4 \ .
\label{chargespectrum}\eeq

\begin{figure}[ht]
\centerline{\epsfxsize=4cm
\epsfbox{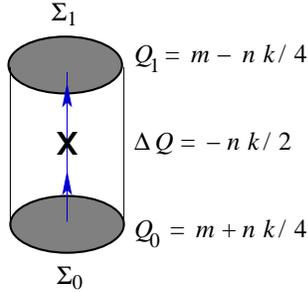}}
\caption{The propagation of a charged particle along a Wilson line
  between left and right
  worldsheets in the presence of a monopole-instanton in the bulk of
  the membrane. The interaction of the particle with the vortex,
  indicated by $\sf X$, shifts its charge from $Q_0$ to
  $Q_1=Q_0+\Delta Q$.}
\label{StringWind}\end{figure}

It is evident from (\ref{chargespectrum}) that the monopole-instantons
shift the spectrum of allowed string momenta, and the magnetic charges
$n\in\zed$ correspond to string winding modes~\cite{CKL1}. As depicted in
Figure~\ref{StringWind}, the presence of monopole-instantons in the
bulk of the membrane, carrying electric charge (\ref{DeltaQVn}), now
shifts the initial charge of a particle from $Q_0$ on the right-moving
worldsheet $\Sigma_0$ to a final charge $Q_1=Q_0+\Delta Q$ on the left-moving
worldsheet $\Sigma_1$. The collection of charges $(Q_0,Q_1)$ live in
the Narain lattice $R\cdot II^{1,1}\subset\real^{1,1}$ with the usual
hyperbolic inner product~\cite{Narain1,NSW1}. When $k=4p/q$ is a
rational number with ${\rm gcd}(p,q)=1$,
the sublattices $Q_0=0$ and $Q_1=0$ are of finite index $pq$ in
$R\cdot II^{1,1}$. Using this inner product and the identification
(\ref{kRrel}) we may compute the spectrum of induced spins
(\ref{anomspins}) as
\bea
2\left(\Delta+\overline{\Delta}\,\right)&=&\frac{2Q_0^2}k+\frac{2Q_1^2}k
\nn\\&=&\frac{2\left(m+\frac{n\,k}4\right)^2}k+\frac{2\left(m-\frac{n\,k}4
\right)^2}k\nn\\&=&m^2\,\frac{\alpha'}{R^2}+n^2\,\frac{R^2}{\alpha'} \
,
\label{spinspectrum}\eea
and the induced angular momentum from propagation between left and
right worldsheets as
\beq
\Delta-\overline{\Delta}=m\,n \ .
\label{angmom}\eeq
The relations (\ref{spinspectrum}) and (\ref{angmom}) reproduce the
mass-shell relation and level-matching condition for bosonic strings
compactified on a circle of radius~$R$. In particular, T-duality in
this picture has the interpretation of interchanging monopole charges
and Dirac charges~\cite{CKS2}.

Because of the relation (\ref{angmom}), any charge non-conserving
process such as that depicted in Figure~\ref{StringWind} must be
accompanied by photon emission in the bulk such that the total angular
momentum of the bulk theory is conserved. With suitable boundary
conditions, and by taking into account all linking
and monopole-instanton processes for charged particle propagation in
the bulk, it is possible to show that all full (non-chiral)
three-dimensional amplitudes for which the initial and final charges
$(Q_0,Q_1)\notin R\cdot II^{1,1}$ vanish identically~\cite{CFKT1}. Thus the
non-perturbative dynamics of the topological membrane naturally
singles out the Narain lattice as the one describing the appropriate
compactified string spectrum. In this way, when $k\in2\rat$, the $U(1)$
topologically massive gauge theory is dual to the $c=1$ conformal
field theory of the extended current algebra of the rational
circle. It is not clear though at this stage how the enhanced,
non-abelian $SU(2)$ Kac-Moody symmetry at level $k=1$ of the conformal
field theory at the self-dual radius $R=\sqrt{\alpha'}$ manifests
itself in the bulk $U(1)$ gauge theory.

Going back to our construction of the heterotic string theory in
Section~\ref{CA}, we now discover a remarkable prediction of
topological membrane theory~\cite{CKL1}. The only way to induce the
required change in spectrum between
the left-movers and right-movers is through non-perturbative charge
non-conserving processes in topologically massive gauge theory, and
these in turn can only be induced by topologically non-trivial gauge
field configurations. Thus the asymmetry between left-moving and
right-moving modes is generated by a {\it compact} gauge group in the
bulk, which translates into a compact target space in the dual string
theory. Thus in the topological membrane construction of the heterotic
string, all spacetime dimensions are required to be compact.

\section{Open Strings\label{OS}}
\setcounter{equation}{0}

We will now show how to describe open strings in terms of topological
membranes, with an eye towards constructing membrane states
corresponding to D-branes. On a first thought, this does not seem
possible to do. The string worldsheet is the boundary of a
three-manifold, $\Sigma=\partial M_3$, and so it is necessarily closed,
$\partial\Sigma=\emptyset$, because the boundary of a boundary is
always empty. This is certainly true for {\it smooth} spaces $M_3$. But
if we allow our membrane geometry to contain singularities, such as
those arising from orbifold constructions, then worldsheet boundaries
can appear at bulk singular points. In this section we will show how
to generate open string worldsheet degrees of freedom by taking
suitable orbifolds of the membrane geometries described in the
previous sections. With the appropriate modifications of these
orbifold operations, everything we say here can also be used to
generate unoriented (Type~I) strings, but we will stick to the open
string description with the ultimate goal of establishing the
appearence of D-branes in topological membrane theory.

\subsection{Worldsheet Orbifolds\label{WO}}

Our construction will be based on the standard description of open
strings as worldsheet orbifolds of closed
strings~\cite{HM1}--\cite{Horava1}. In conformal field
theory, a correlation function on a worldsheet $\Sigma$ is completely
determined by a choice of vector in the space of conformal blocks
associated to a double cover $\widehat{\Sigma}$ of $\Sigma$, obeying
factorization constraints and modular invariance~\cite{CL1}. The double
$\widehat{\Sigma}$ is a closed oriented Riemann surface which
generates the worldsheet $\Sigma$ via the orbifold
\beq
\Sigma~=~\widehat{\Sigma}~/~\sigma
\label{Sigmadoublegen}\eeq
with respect to an anti-conformal involution
$\sigma:\widehat{\Sigma}\to\widehat{\Sigma}$, $\sigma\circ\sigma={\rm
  id}$ which generates the worldsheet parity symmetry group
$\zed_2^{\rm ws}$. The fixed points of $\widehat{\Sigma}$ under the
involution $\sigma$ correspond to boundary points of $\Sigma$. For
example, if $\Sigma$ is a disk ${\mathbb D}^2$, then its double
$\widehat{\Sigma}$ is a sphere $\sphere^2$ and
$\sigma:\sphere^2\to\sphere^2$ is reflection about the equatorial
plane. If the worldsheet $\Sigma$ is oriented and closed, then its
double $\widehat{\Sigma}=\Sigma\amalg-\Sigma=\Sigma_0\amalg\Sigma_1$
is also oriented and closed, and supports a full {\it non-chiral}
conformal field theory.

Generally, the worldsheet orbifold group $G_{\rm orb}$ combines $\zed_2^{\rm
  ws}$ together with the target space symmetry group $G$ such
  that $G_{\rm orb}\subset G\times\zed_2^{\rm ws}$. We will usually deal
  with ``standard'' worldsheet orbifolds in which it is a direct
  product $G_{\rm orb}=G_0\times\zed_2^{\rm ws}$, $G_0\subset G$. An
  open string may be viewed as an orbifold
  $\mathcal{O}_{\rm str}=\sphere^1/\zed_2$ of a closed string, with the cyclic
  group $\zed_2$ generated by the reflection $\sphere^1\to\sphere^1$
  through the equatorial line of the circle. Its fundamental group is
  the infinite dihedral group
\beq
\pi_1(\mathcal{O}_{\rm str})=
\mathfrak{D}_\infty=\zed_2\star\zed_2=\zed_2\ltimes\zed_2 \ ,
\label{dihedral}\eeq
where $\star$ denotes the free product of discrete abelian groups.

The monodromies of fields in the open string sector $\mathcal{O}_{\rm str}$
correspond to the representations of (\ref{dihedral}) in the orbifold
group such that the triangle diagram
\bea
\zed_2\star\zed_2&~\longrightarrow~&G_{\rm orb}\nn\\
\searrow& &\swarrow\nn\\ &\zed_2^{\rm ws}&
\label{commtriangle}\eea
is commutative. The partition function of the corresponding open
worldsheet field theory is then a sum over all such monodromies of the
form
\beq
Z_\Sigma(\mbf m)=\frac1{|G_{\rm orb}|^g}~\sum_{\alpha\,:\,\pi_1(\Sigma)\to
G_{\rm orb}}Z_\Sigma(\alpha;\mbf m) \ .
\label{wsorbpartfn}\eeq
For example, with this prescription the cylinder amplitude takes the
form
\beq
Z_{\real\times\sphere^1}(t)=\frac1{|G_{\rm orb}|}~\sum_{\stackrel{\scriptstyle
g_1,g_2,h}{\scriptstyle g_i^2=\id\,,\,[g_i,h]=\id}}
Z_{\real\times\sphere^1}(g_1,g_2,h;t) \ ,
\label{cylampl}\eeq
where the modulus $t\in\real$ of the cylinder corresponds to the
circumference of the non-contractible cycle of $\real\times\sphere^1$,
and the monodromy elements in the sum (\ref{cylampl}) assume the
standard forms $g_i=(\tilde g_i,\sigma)$, $h=(\tilde h,{\rm id})$ with
$\tilde g_i,h\in G_0\subset G$.

\subsection{Membrane Orbifolds\label{MO}}

\begin{figure}[ht]
\centerline{\epsfxsize=10cm
\epsfbox{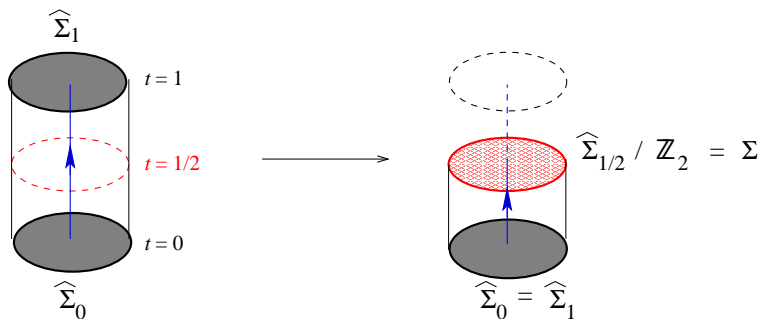}}
\caption{The orbifold of the topological membrane creates an open
  string worldsheet $\Sigma$ at the time reversal fixed point
  $t=\frac12$ and identifies the holomorphic and antiholomorphic
  sectors $\widehat{\Sigma}_0=\widehat{\Sigma}_1$ of the associated
  closed string double.}
\label{TMOrbifold}\end{figure}

We will now cast the orbifold constructions of the previous subsection
into the framework of topological
membranes~\cite{FRS1,FFFS1,bb,CFKS1,Horava2,CFK1}. The {\it chiral}
sector is induced by introducing the connecting three-manifold
\beq
M_\Sigma~=~\left(\widehat{\Sigma}\times[0,1]\right)~/~\zed_2 \ ,
\label{conn3man}\eeq
where the cyclic group $\zed_2$ combines the worldsheet involution
$\sigma$ on the double $\widehat{\Sigma}$ with time reversal ${\sf
  T}:t\mapsto1-t$ on the interval $[0,1]$. This is a three-manifold
with boundary $\partial M_\Sigma=\widehat{\Sigma}$. For example, for
$\Sigma={\mathbb D}^2$, $\widehat{\Sigma}=\sphere^2$, the connecting
three-manifold is the ball $M_\Sigma={\mathbb D}^3$ with boundary
$\partial{\mathbb D}^3=\sphere^2$. However, for the construction of
orbifold membrane amplitudes to be carried out in subsequent sections,
it is more convenient to use a slightly
different version of the $\zed_2$ orbifold in
(\ref{conn3man})~\cite{bb,CFKS1,Horava2,CFK1}, in
which the time reversal involution $\sf T$ acts only to create a new
open surface at its fixed point $t=\frac12$ on $[0,1]$. This operation
is depicted in Figure~\ref{TMOrbifold}.

Let us consider the allowed orbifold operations on the membrane
which are discrete symmetries of topologically massive gauge
theory defined on the three-geometry
$\widehat{\Sigma}\times[0,1]$~\cite{CFK1}. The Chern-Simons action is odd under
both time-reversal $\sf T$ and the standard worldsheet parity $\sigma=\sf
P$. Because of the presence of the additional Yang-Mills kinetic term in the
topologically massive gauge theory action, it is not difficult to see
that the only discrete three-dimensional spacetime symmetries
compatible with bulk gauge invariance are the combinations $\sf PT$ and $\sf
PCT$, where $\sf C$ is the usual charge conjugation operation in three
dimensions. These are therefore the only $\zed_2$ actions that we are
allowed to take in defining our gauge theory on the orbifold
(\ref{conn3man}). The action functional on the connecting three-manifold is
then
defined in terms of the previous gauge theory action as
\beq
2S_{\rm TMGT}^{M_\Sigma}\bigl[A
\bigr]=S_{\rm TMGT}^{\widehat{\Sigma}\times[0,1]}
\bigl[\,\widehat{A}\,\bigr] \ ,
\label{STMGTorb}\eeq
where $\widehat{A}$ is the extension of the gauge field $A$ from
$M_\Sigma$ to the covering cylinder $\widehat{\Sigma}\times[0,1]$.

For an induced string theory which is compactified on a circle as in
the previous section, it is also straightforward to work out how the
allowed $\sf PT$ and $\sf PCT$ automorphisms of the topological
membrane act on the charge spectrum (\ref{chargespectrum}) and
appropriately truncate it according to the standard boundary
conditions in open string theory~\cite{CFK1}. For the $\sf PT$
orbifold of topologically massive gauge theory one finds
\beq
{\sf PT}\,:\,Q~\longmapsto~-Q \ .
\label{PTQ}\eeq
After performing the orbifold operation which identifies left and
right worldsheets, the identification $Q_1=-Q_0$ leaves a spectrum of
pure winding modes $Q=n\,k/4$ and corresponds to Dirichlet boundary
conditions on the open string embedding fields. Similarly, one has
\beq
{\sf PCT}\,:\,Q~\longmapsto~Q
\label{PCTQ}\eeq
and the orbifold identification $Q_1=Q_0$ truncates the charge
spectrum to pure KK-modes $Q=m$, corresponding to Neumann boundary
conditions on the open strings.

\subsection{Singleton Orbifolds\label{SO}}

The final truncation to open strings that we need to make is the
appropriate specification of how the harmonic modes of the gauge
connections map under the orbifold operations in
(\ref{conn3man})~\cite{FFFS1}. The worldsheet involution
$\sigma:\widehat{\Sigma}\to\widehat{\Sigma}$ induces an involutive
isomorphism on homology $\sigma_*:{\rm
  H}_1(\,\widehat{\Sigma},\real)\to{\rm
  H}_1(\,\widehat{\Sigma},\real)$. The homology group ${\rm
  H}_1(\,\widehat{\Sigma},\real)$ is a symplectic vector space over
$\real$ with respect to the canonical intersection form on the closed
oriented Riemann surface $\widehat{\Sigma}$. The subspaces
$\mathrm{L}_\pm(\,\widehat{\Sigma}\,)\subset{\rm
  H}_1(\,\widehat{\Sigma},\real)$ defined as the $\pm\,1$ eigenspaces of
the involution, $\sigma_*\mathrm{L}_\pm(\,\widehat{\Sigma}\,)
=\pm\,\mathrm{L}_\pm(\,\widehat{\Sigma}\,)$, are Lagrangian subspaces
with respect to this symplectic structure and the homology group
decomposes into the orthogonal direct sum
\beq
{\rm H}_1\bigl(\,\widehat{\Sigma}\,,\,\real\bigr)=
\mathrm{L}_+\bigl(\,\widehat{\Sigma}\,\bigr)\oplus
\mathrm{L}_-\bigl(\,\widehat{\Sigma}\,\bigr) \ .
\label{H1Lagrsubsp}\eeq
We then identify the homology of the worldsheet orbifold obtained from
the double $\widehat{\Sigma}$ as the Lagrangian subspace
\beq
{\rm H}_1\bigl(\Sigma\,,\,\real\bigr)=\mathrm{L}_-\bigl(\,\widehat{\Sigma}
\,\bigr) \ .
\label{H1orb}\eeq

The identification (\ref{H1orb}) has a very natural geometrical
interpretation from the topological membrane perspective. Given the
connecting three-manifold (\ref{conn3man}), the canonical inclusion
$\imath:\widehat{\Sigma}=\partial M_\Sigma\to M_\Sigma$ induces a
homomorphism $\imath_*:{\rm H}_1(\,\widehat{\Sigma},\real)\to{\rm
  H}_1(M_\Sigma,\real)$ with null space
\beq
\ker\bigl(\imath_*\bigr)=\mathrm{L}_-\bigl(\,\widehat{\Sigma}\,\bigr) \ .
\label{kerimath}\eeq
In other words, the homology group (\ref{H1orb}) of the open string
worldsheet $\Sigma$ consists of those homology cycles of its double
$\widehat{\Sigma}$ which are contractible in the corresponding
connecting three-manifold $M_\Sigma$. These truncated topological
degrees of freedom can also be related more directly to the membrane geometry
$\widehat{\Sigma}\times[0,1]$ by using a $\zed_2$-equivariant version
of this homological construction~\cite{bb}. Equivalently, since $M_\Sigma$
retracts to $\Sigma=\widehat{\Sigma}/\zed_2$, we can pick an arbitrary
imbedding $\jmath:\Sigma\hookrightarrow M_\Sigma$ to carry out the
appropriate truncation~\cite{FRS1,FFFS1}.

\section{Orbifold Amplitudes\label{OrbAmpl}}
\setcounter{equation}{0}

The remainder of this article is devoted to explaining how to use the
set-up of the previous section to describe states corresponding to
D-branes in three-dimensional terms. In this section we will describe
how open string amplitudes easily arise from orbifolds of
topologically massive gauge theory, and how they can be used to
systematically derive open string vertex operators in the induced
two-dimensional boundary conformal field theory. We shall see that the
vertex operators for both Neumann and Dirichlet branes arise very
naturally in this formalism.

\subsection{Open String Amplitudes\label{OSMA}}

We will begin with a heuristic explanation of how three-dimensional
orbifolds of the previously derived closed string amplitudes of
topologically massive gauge theory naturally induce the expected open
string amplitudes~\cite{CFKS1}. Recall the membrane inner product
(\ref{statsumTMGT}) which we write in the form
\beq
Z_{\widehat{\Sigma}}^{\rm TMGT}=\sum_{\lambda,\lambda'
\in(\Lambda_{G_k}^*/\Lambda^{~}_{G_k})^g}
\zeta^{\lambda\lambda'}~\psi^{~}_\lambda(0)\otimes
\overline{\psi}^{~}_{\lambda'}(1) \ ,
\label{Zclosedinner}\eeq
where the non-negative integers $\zeta^{\lambda\lambda'}$ ensure
modular invariance of the worldsheet amplitude. We have emphasized its
interpretation as the propagation amplitude for the evolution of an initial
state $\psi^{~}_\lambda(0)$, inserted at time $t=0$ on the holomorphic
boundary of the membrane geometry depicted on the left-hand side of
Figure~\ref{TMOrbifold}, through the bulk of the membrane to a final
state $\psi^{~}_{\lambda'}(1)$ at time $t=1$ on the anti-holomorphic
sector of the closed string double~$\widehat{\Sigma}$. It is regarded
as a vector in the Chern-Simons Hilbert space
\beq
Z_{\widehat{\Sigma}}^{\rm TMGT}~\in~
{\mathcal H}_{\widehat\Sigma}^{\rm CS}\otimes
\overline{{\mathcal H}_{\widehat\Sigma}^{\rm CS}}
\label{closedCSHilbert}\eeq
of non-chiral conformal blocks.

We can use this bilinear form to induce an amplitude corresponding to
propagation of states between times $t=0$ and $t=\frac12$ in the
membrane geometry depicted on the right-hand side of
Figure~\ref{TMOrbifold} by taking a membrane orbifold. For this, we
insert a complete set of states $\psi^{~}_\rho(\frac12)$ into
(\ref{Zclosedinner}) at $t=\frac12$ to write
\bea
Z_{\widehat{\Sigma}}^{\rm TMGT}&=&\sum_{\lambda,\lambda'
\in(\Lambda_{G_k}^*/\Lambda^{~}_{G_k})^g}
\zeta^{\lambda\lambda'}~\psi^{~}_\lambda\bigl(0\bigr)\nn\\&&\otimes~
\sum_{\rho\in(\Lambda_{G_k}^*/\Lambda^{~}_{G_k})^g}\overline{\psi}^{~}_\rho
\bigl(\mbox{$\frac12$}
\bigr)~\psi^{~}_\rho\bigl(\mbox{$\frac12$}\bigr)~\otimes~
\overline{\psi}^{~}_{\lambda'}\bigl(1\bigr) \ .
\label{Zcompleteinsert}\eeq
Let us now consider the orbifold involutions of Section~\ref{MO}. At
the orbifold fixed point $t=\frac12$, Wilson lines transform as
$W_{C(\rho)}[A]\mapsto W_{-C(\rho)}[A]$ under both the $\sf PT$
and $\sf PCT$ orbifolds whose overall effect is to change the orientation
of the worldline $C$ of the charge $\rho$. Making the desired
orbifold identifications thereby leads to the requirement
\beq
W_{C(\rho)}[A]=W_{-C(\rho)}[A]
\label{Wilsonorbcond}\eeq
for arbitrary oriented contours $C\subset M_3$. If
$C\subset\widehat{\Sigma}_{1/2}$, then this is only possible in the
trivial charge sector $\rho=0$. Since the conformal field theory
wavefunctions are generated by Wilson lines lying entirely in
$\widehat{\Sigma}$ (via the appropriate monopole-instanton processes),
it follows that the only membrane state possible at
$\widehat{\Sigma}_{1/2}$ is the identity character state
$\psi^{~}_{\rho=0}(\frac12)=1$.

Thus, after orbifolding, the insertion of a complete set of states in
(\ref{Zcompleteinsert}) leaves simply the orthogonal projection,
acting on the Chern-Simons Hilbert space, onto those membrane states which are
invariant under interchange of left and right moving worldsheet
modes. We thereby find the orbifold amplitude
\beq
Z_{\Sigma}^{\rm orb}=\sum_{\lambda,\lambda'\in[
(\Lambda_{G_k}^*/\Lambda^{~}_{G_k})^g]^{~}_{\rm orb}}
\zeta^{\lambda\lambda'}~\psi^{~}_\lambda(0)\otimes
\mbox{$\frac12$}\,\left(\id+\overline{\sigma^{~}_\#}\,\right)
\overline{\psi}^{~}_{\lambda'}(1) \ ,
\label{ZSigmaorbproj}\eeq
where $\sigma^{~}_\#:{\mathcal H}_{\widehat\Sigma}^{\rm
  CS}\to{\mathcal H}_{\widehat\Sigma}^{\rm CS}$ is the induced
  representation of the $\sf PT$ or $\sf PCT$ involution
  on the three-dimensional states of the membrane and the subscript
  orb denotes the appropriate truncation of the charge lattice under
  the orbifold operation. After taking into
  account the induced identifications
  $\psi^{~}_\lambda(0)=\psi^{~}_\lambda(1)=\psi_\lambda^{\rm orb}$,
  the amplitude (\ref{ZSigmaorbproj}) acquires the form~\cite{CFKS1}
\beq
Z_{\Sigma}^{\rm orb}=\sum_{\lambda\in[
(\Lambda_{G_k}^*/\Lambda^{~}_{G_k})^g]^{~}_{\rm orb}}
h^\lambda~\psi_\lambda^{\rm orb}
\label{ZSigmaorbfinal}\eeq
and it is regarded as a vector in the Hilbert space
\beq
Z_{\Sigma}^{\rm orb}~\in~{\mathcal H}_{\widehat\Sigma}^{\rm CS} \ .
\label{openCSHilbert}\eeq

Thus we easily derive the fact that for closed oriented worldsheets
$\widehat{\Sigma}$, the partition function is a bilinear form in the
characters of the induced conformal field theory, while for open (or
unoriented) worldsheets $\Sigma$ it is simply linear in the
characters~\cite{CFKS1,Horava2}. Moreover, these same arguments serve
to show that open string membrane amplitudes can be regarded as square
roots of the corresponding closed string amplitudes~\cite{bb}
\beq
\left|\left\langle\Psi_0\,\left|\,\Psi_{1/2}\right.\right\rangle_{\rm
    orb}\right|=
\sqrt{\left|\bigl\langle\Psi_0\,\bigl|\,\Psi_1\bigr.\bigr\rangle
\right|} \ .
\label{openclosedamplrel}\eeq
In other words, closed strings come in a double volume of open
strings.

\subsection{Brane Vertex Operators\label{BVO}}

We can also carry out the construction of orbifold amplitudes more
precisely, using our previous path integral formalism for membrane
amplitudes and the appropriate Schr\"odinger wavefunctionals~\cite{bb}. For
this, we will consider the simplest instance of $U(1)$ topologically
massive gauge theory minimally coupled to a conserved current
represented by a (dual) one-form $J=\mbf J+J_0~\dd t$ on
$M_3=\widehat{\Sigma}\times[0,1]$. The action is
\beq
S_J[A]=S_{\rm TMGT}[A]+\int_{M_3}A\wedge*J
\label{SJAdef}\eeq
while the continuity equation for the source is given by
\beq
\dd*J=0 \ .
\label{conteqn}\eeq
The appropriate boundary conditions which are compatible with the $\sf
PT$ and $\sf PCT$ orbifold involutions of the membrane require us to
fix the sources in terms of an auxilliary one-form
$\tilde{Y}=\tilde Y^{1,0}+\tilde Y^{0,1}$ on the closed string
double $\widehat{\Sigma}$ as
\bea
J_0~\dd{\rm vol}_{\widehat{\Sigma}}&=&\dd\tilde Y \ , \nn\\
J^{1,0}&=&\mbox{$\frac\mu2$}\,*_2\tilde Y^{0,1} \ , \nn\\
J^{0,1}&=&\mbox{$\frac\mu2$}\,*_2\tilde Y^{1,0} \ ,
\label{sourcefix}\eea
with $\mu$ the topological photon mass (\ref{topmassTMGT}). This
restriction further guarantees that the amplitudes, defined originally in
terms of wavefunctionals living on the boundaries $\widehat{\Sigma}_0$
and $\widehat{\Sigma}_1$, can be systematically extended into the bulk
of the membrane. It is also the key to constructing vacuum
wavefunctionals of the matter-coupled gauge theory which respect gauge
invariance as before~\cite{bb}.

It is now straightforward to incorporate the addition of charged
matter in this way for {\it arbitrary} source configurations to the
construction of vacuum Schr\"odinger wavefunctionals, extending the
constructions we carried out earlier for charged particle Wilson
lines. Doing so, and then carefully computing the orbifold of the
corresponding membrane inner product, we arrive at the orbifold
partition function~\cite{bb}
\beq
Z_{\Sigma}^{\rm orb}=\left\langle\Psi_0^{\rm vac}\,\left|\,
\Psi_{1/2}^{\rm vac}\right.\right\rangle_{\rm orb}
\label{orbpartfndef}\eeq
with the vacuum wavefunctionals
\bea
\Psi_0^{\rm vac}\left[\mbf A\,,\,\tilde Y\right]&=&
\e^{-\int_{\widehat{\Sigma}}A^{1,0}\wedge(\frac{|k|}{4\pi}\,A^{0,1}+
\tilde Y^{0,1})}\nn\\&&\times\,\int\DD\phi~
\e^{-\frac{|k|}{4\pi}\,\int_{\widehat{\Sigma}}(\dd\phi\wedge*_2\,
\dd\phi-2\,\partial\phi\wedge A^{0,1})} \ , \label{Psiorb0vac}
\\\Psi_{1/2}^{\rm vac}\left[\mbf A\,,\,\tilde Y\right]&=&
\int\DD\phi~\e^{\int_{\partial\Sigma}\phi\,(\tilde Y^\parallel-
\frac{|k|}{8\pi}\,A^\parallel)}\nn\\&&\times\,
\e^{-\frac{|k|}{8\pi}\,\int_\Sigma[A^{1,0}\wedge(\,\overline{\partial}
\phi-\frac{8\pi}{|k|}\,\tilde Y^{0,1})-A^{0,1}\wedge(
\partial\phi-\frac{8\pi}{|k|}\,\tilde Y^{1,0})]} \ .
\label{Psiorb12vac}\eea
Here $\phi$ is the boson field of the induced $c=1$ conformal field
theory in this case, and in (\ref{Psiorb12vac}) the superscript
$\parallel$ denotes the projections of the one-forms onto the
cotangent bundle over the boundary $\partial\Sigma$ in the open string
worldsheet $\Sigma$. A simple field redefinition in the path integrals
removes the source terms from the bulk of $\Sigma$, and thus in this
case a change of conformal background due to the insertion of bulk
charged matter is induced solely by boundary deformations.

Let us now insert the local Hodge decomposition on $\widehat{\Sigma}$
of the charged matter deformation given by
\beq
\tilde Y=\frac{|k|}{4\pi}\,\bigl(\dd Y_{\rm D}+*_2\,\dd Y_{\rm
  N}\bigr)
\label{tildeYHodge}\eeq
with $Y_{\rm D},Y_{\rm N}\in\Omega^0(\,\widehat{\Sigma}\,)$. The
absence of harmonic modes in (\ref{tildeYHodge}) is imposed by the
requirement that the sources be non-dynamical in $M_3$. In
addition, we use the Hodge decomposition (\ref{Hodgedecompabelian}) of the
topologically massive gauge field $\mbf A$ on the closed string double
$\widehat{\Sigma}$. The functional integrations over the harmonic
degrees of freedom $\mbf a$ in
(\ref{orbpartfndef},\ref{Psiorb12vac}) for the $\sf PT$ and $\sf
PCT$ orbifolds of the topological membrane then respectively yield the
delta-function constraints~\cite{bb}
\bea
{\sf PT}~~&:&~~\delta\left(\,\nabla^2(\phi-2\,Y_{\rm
    D})\bigm|_\Sigma\,\right)~\delta\left(\,(\phi-Y_{\rm
    D})\bigm|_{\partial\Sigma}\,\right) \ , \label{PTconstr}\\
{\sf PCT}~~&:&~~\delta\left(\,\nabla^2(\phi-2\,Y_{\rm
    D})\bigm|_\Sigma\,\right)~\delta\left(\,i_{\partial_\perp}\,\dd
\phi\bigm|_{\partial\Sigma}\,\right) \ .
\label{PCTconstr}\eea
The first delta-function in both (\ref{PTconstr}) and
(\ref{PCTconstr}) imposes (after a simple field redefinition) the bulk
equation of motion for the {\it free} scalar field $\phi$ on the open
string worldsheet $\Sigma$. For the $\sf PT$ orbifold the second
delta-function imposes Dirichlet boundary conditions for $\phi$ on
$\partial\Sigma$, while for the $\sf PCT$ orbifold it selects Neumann
boundary conditions. Thus the only roles played by the wavefunctional
(\ref{Psiorb12vac}) at the orbifold branch point are to enforce the
worldsheet equations of motion and to select the appropriate boundary
conditions of open string theory. Otherwise they simply correspond to
inserting the identity character state into the membrane inner
product, consistently with what we argued in the previous subsection.

What is particular interesting about the orbifold vacuum
wavefunctional (\ref{Psiorb12vac}) is the boundary exponential
term. After functional integration, for the $\sf PT$ and $\sf PCT$
orbifolds of the topological membrane it yields a boundary deformation
of the usual bulk $\sigma$-model action given respectively by~\cite{bb}
\bea
{\sf PT}~~&:&~~V_{\rm D}=\e^{-\frac{|k|}{4\pi}\,\int_{\partial\Sigma}
Y_{\rm D}~i_{\partial_\perp}\,\dd\phi} \ , \label{DBraneVertex}\\
{\sf PCT}~~&:&~~V_{\rm N}=\e^{-\frac{\ii|k|}{4\pi}\,\int_{\partial\Sigma}
Y_{\rm N}~i_{\partial_\parallel}\,\dd\phi} \ .
\label{PhotonVertex}\eea
The insertion (\ref{DBraneVertex}) into the conformal field theory
path integral is the vertex operator for a D-brane described by the
collective coordinate $Y_{\rm D}$ in the target space direction
$\phi$. Using the identification (\ref{kRrel}), one sees that this
correspondence is exact. Similarly, the operator (\ref{PhotonVertex})
is the open string photon Wilson line for the $U(1)$ gauge field
component $Y_{\rm N}/2\pi\,\alpha'$ in direction~$\phi$. Thus D-branes in
topological membranes correspond to charged matter on an orbifold line
in three dimensions. In this picture, the collective coordinate of the
D-brane is controlled by the bulk charge distribution. This is in
perfect harmony with our description of deformed conformal field
theories through the addition of charged matter fields in the bulk
that was given in Section~\ref{DCFT}.

\section{Boundary States\label{BS}}
\setcounter{equation}{0}

The results of Section~\ref{BVO} provide very encouraging
evidence that the wavefunctionals (\ref{Psiorb12vac}) describe
three-dimensional states corresponding to D-branes. Of course, the
fact that they induce the correct vertex operators in the effective
$\sigma$-model action is not a proof of this fact, and one now needs
to carefully explore to what extent the dynamics of the topological membrane
truly captures the physics of D-branes. One way to proceed towards
this goal is to analyse to what extent the orbifolds of topologically
massive gauge theory induce {\it boundary} conformal field
theories. In this section we will show how to construct standard
closed string boundary states in three-dimensional terms and also how
the bulk-boundary correspondence of conformal field theory manifests
itself in the topological membrane approach.

\subsection{Ishibashi States\label{IS}}

Let us recall the vacuum wavefunctionals
$\Xi^{\{\lambda_i\}}[\{z_i\};\mbf A]$ in (\ref{Xilambdaidef})
describing states of chiral, non-dynamical charged particles in the topological
membrane. They are elements of the complex vector space $\mbf V$ as
given in (\ref{Xirepspaces}), and they correspond to primary
chiral field insertions on the closed Riemann surface
$\widehat{\Sigma}$. Later on, we will see how to also incorporate
string descendent fields into the membrane wavefunctionals and
amplitudes. But for the time-being, we assume that this has been done
and extend the representation spaces $V_i$ to the appropriate
Virasoro-Sugarawa modules by application of the corresponding Virasoro
descendent fields. We will also denote these modules by $V_i$.

We shall now analyse the structure of these operators in the case
$n=1$, corresponding to a single particle insertion of charge
$\lambda$. Consider a vertical Wilson line operator $W^\lambda[A]$ in
the instance when the bulk topologically massive gauge theory is
assumed to possess its full discrete $\sf PCT$ invariance. Then the
orbifold involution acts on this Wilson line as
\beq
{\sf PCT}\,:\,W^\lambda[A]~\longmapsto~W^\lambda[A]
\label{PCTWlambdaA}\eeq
and thus the charge non-conserving monopole-instanton induced
processes are suppressed. Moreover, the corresponding wavefunctional
(\ref{Xilambdaidef}) is an operator
\beq
\Xi^\lambda[z;\mbf A]\,:\,V_\lambda~\longrightarrow~V_\lambda \ .
\label{Xilambdaop}\eeq

By gluing together left and right moving worldsheet sectors as before,
one finds that the $\sf PCT$-invariant membrane states describing
propagation between the boundaries $\widehat{\Sigma}_0$ and
$\widehat{\Sigma}_1$ are given by the vacuum wavefunctionals
\beq
\Phi^\lambda[z,\overline{z}\,;\mbf A]=\Xi^\lambda[z;\mbf A]\otimes
\overline{\Xi^{\bar\lambda=\lambda}[\,\overline{z}\,;\mbf A]} \ .
\label{PCTinvstates}\eeq
The identification of left and right moving charges
$\bar\lambda=\lambda$ arises from the identification under worldsheet
parity ${\sf P}:\widehat{\Sigma}_0\equiv\widehat{\Sigma}_1$. The
operator (\ref{PCTinvstates}) acts only in the diagonal, left-right
symmetric product $V^{~}_\lambda\otimes V_{\bar\lambda=\lambda}\subset\mbf
V$. It is therefore proportional to the orthogonal projection
\beq
P_\lambda=\sum_{I\in\mathcal{I}}\,\bigl|\lambda\,,\,I\bigr\rangle\bigl
\langle\,\bar\lambda=\lambda\,,\,I\bigr|
\label{Plambdaproj}\eeq
onto this subspace.

By choosing the proportionality constant to be $1$, and using the
natural inner product on the Hilbert space $V_{\bar\lambda}$, the
homomorphism $\Phi^\lambda:V_{\bar\lambda}\to V^{~}_\lambda$ is in a
one-to-one correspondence with the closed string Ishibashi
state~\cite{Ish1,BPPZ1} given by
\beq
|\lambda\rangle\!\rangle^{\rm D}=\sum_{I\in\mathcal{I}}\,|\lambda,I
\rangle\otimes U^{~}_{\sf P}\,U^{~}_{\sf C}|\lambda,I\rangle \ ,
\label{DIshibashistate}\eeq
where the anti-unitary operators $U^{~}_{\sf P}$ and $U^{~}_{\sf C}$ implement
the actions of the parity and charge conjugation automorphisms on the
right-moving
Hilbert space. The vector (\ref{DIshibashistate}) is just a Dirichlet
boundary state of the induced closed string theory. In a completely
analogous manner, by assuming only the $\sf PT$ sub-invariance of the
bulk quantum field theory, one can readily construct Neumann Ishibashi
boundary states~\cite{bb}. In this case one must account for non-trivial
monopole-instanton transitions in the bulk. Again it is not clear
though how, for the free boson at the self-dual radius, the family of
branes parametrized by the $SU(2)$ group manifold at level
$k=1$~\cite{GRW1} appears here, with the Dirichlet brane corresponding
to the identity element of $SU(2)$. A possible clue may lie in
boundary state of~\cite{CJrKLM1} which is a global $SU(2)$ rotation of
the usual Neumann boundary state.

\subsection{The Bulk-Boundary Correspondence\label{TB-BC}}

Let us now see how the usual bulk-boundary correspondence of
two-dimensional conformal field theory manifests itself in the three
dimensional framework. According to this principle, the $n$-point
correlators on an open surface $\Sigma$ are in one-to-one
correspondence with chiral $2n$-point correlation functions on the
double $\widehat{\Sigma}$~\cite{CL1}. The interaction of a local field with
$\partial\Sigma$, in the form of boundary conditions, is then
simulated by the interaction between mirror images of the same
holomorphic field on $\widehat{\Sigma}$, carrying conjugate primary
charges $\lambda,\bar\lambda$.

\begin{figure}[ht]
\centerline{\epsfxsize=10cm
\epsfbox{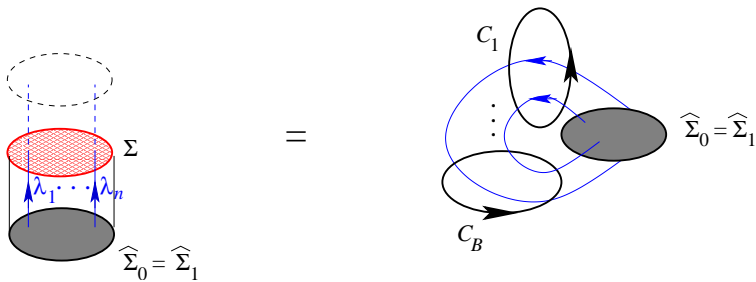}}
\caption{Three-dimensional representation of the bulk-boundary computation of
  $n$-point conformal correlation functions. The interactions of
  local field insertions with $\partial\Sigma$ are represented by
  the interactions with Wilson loops $W^{\lambda_\alpha}[A]$
  corresponding to the open string boundaries $C_\alpha$. The chiral
  correlator on the right is computed using the finite temperature
  prescription of Section~\ref{PLTMGT} along with the linking rules
  described in Section~\ref{VF}.}
\label{Wilsonintbdry}\end{figure}

To each conformal boundary condition $\alpha$, we insert
a Wilson loop $W^{\lambda_\alpha}[A]$ in the bulk corresponding to a
prescribed representation $\lambda_\alpha$ of the gauge group. More
precisely, let us suppose that the boundary of the open string
worldsheet consists of $B$ connected components given by the disjoint
union
\beq
\partial\Sigma=\coprod_{\alpha=1}^BC_\alpha \ .
\label{disjunion}\eeq
In the double $\widehat{\Sigma}$, the pre-image of each loop
$C_\alpha$, $\alpha=1,\dots,B$ is a $\zed_2$-invariant equatorial
circle corresponding to a Wilson loop in the covering cylinder
$\widehat{\Sigma}\times[0,1]$, which becomes a circle of singular
points in the three-dimensional
orbifold~(Figure~\ref{Wilsonintbdry}). Any connected component
$C_\alpha$ of the singular locus of the orbifold $\Sigma$ can be
represented as a sum over Wilson loops with the topology of
$C_\alpha$~\cite{FFFS1,Horava2}.

These facts can all be derived systematically by examining orbifold
membrane amplitudes with the prescribed boundary conditions on
$\partial\Sigma$~\cite{bb}. The boundary conditions are charges of external
matter inside the topological membrane, and are thereby naturally
labelled by primary charges, just as in conformal field theory. These
are the boundary conditions which preserve all bulk symmetries. This
three-dimensional prescription thus provides a very effective
computation of the correlation functions in boundary conformal field
theory. An open problem in this context is the description of
symmetry-breaking boundary conditions in the three-dimensional
framework, and hence of D-branes which are not maximally symmetric.

\section{The Cardy Condition\label{TCC}}
\setcounter{equation}{0}

We can now describe one of the fundamental results of boundary
conformal field theory, the celebrated Cardy condition. Just like in
the case of the Verlinde formula, it has a very natural dynamical
origin within the framework of open topological membranes. This
analysis allows one to select the totality of three-dimensional states
corresponding to D-branes based on the single guiding principle of
bulk gauge invariance.

\subsection{Fundamental Brane States\label{FBS}}

The Cardy condition of boundary conformal field theory~\cite{Cardy1}
follows from the equality of the closed string cylinder amplitude,
computed as a
matrix element between two boundary states, and the open string
annulus amplitude with the corresponding boundary conditions. In the
language of topological membranes, it arises as a compatibility
condition between conservation of orbifold charges and bulk gauge invariance in
$M_3$~\cite{bb}. To understand this point, let us consider the ``fundamental''
wavefunctionals
\beq
\Upsilon^\lambda[z,\overline{z}\,;\mbf A]=
\sum_{\lambda'\in(\Lambda_{G_k}^*/\Lambda^{~}_{G_k})^g}
\beta_{\lambda'}^{~~\lambda}~\Phi^{\lambda'}[z,\overline{z}\,;\mbf A]
\label{fundwavefns}\eeq
given as linear superpositions of the Ishibashi wavefunctionals
(\ref{PCTinvstates}) describing D-brane states. The coupling
coefficients may be fixed by demanding that membrane inner products
with the wavefunctionals (\ref{fundwavefns}) be determined as overlaps
with the trivial $\lambda=0$ Wilson line (corresponding to the
identity character state as described earlier) in the orbifold
topologically massive gauge theory. An elementary computation
gives~\cite{bb}
\beq
\beta_{\lambda'}^{~~\lambda}=\frac{{\sf S}_{\lambda\lambda'}}
{{\sf S}_{\lambda0}} \ ,
\label{betaCardysol}\eeq
and thereby yields a remarkably simple derivation of the Cardy solution (up to
normalization) of the sewing constraints in boundary conformal field
theory~\cite{Cardy1}.

The coupling coefficients (\ref{betaCardysol}) are described
dynamically by the Hopf linking amplitudes (\ref{SmatrixHopf}). With
them we may compute the matrix elements~\cite{bb}
\beq
\bigl\langle1\,\bigl|\,W^{\lambda'}\Upsilon^\lambda\bigr.\bigr\rangle=
{\sf S}_{-\lambda,\lambda'}\,\bigl\langle1\,\bigl|\,\Upsilon^\lambda\bigr.
\bigr\rangle \ .
\label{propmatrixelts}\eeq
We interpret (\ref{propmatrixelts}) to mean that as a charged particle
propagates through the bulk it interacts with a soliton-like defect
described by the boundary state
(\ref{fundwavefns},\ref{betaCardysol}), producing the usual Hopf
linking factors in $\sphere^3$. In the worldsheet picture, this defect
clearly corresponds to a D-brane, producing the correct charge
deformation of the boundary field.

\subsection{Surgery Calculation\label{SP}}

\begin{figure}[ht]
\centerline{\epsfxsize=8cm
\epsfbox{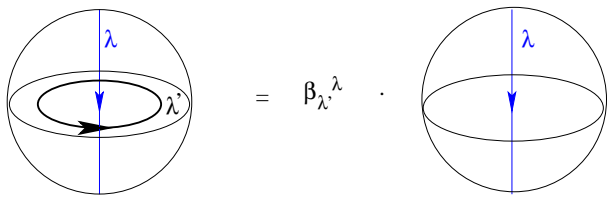}}
\caption{Three-dimensional version of the representation of the
  correlator of a bulk field on the disk in terms of the standard
  two-point conformal block on the sphere.}
\label{1ptfndisk}\end{figure}

The relationship (\ref{fundwavefns},\ref{betaCardysol}) between brane
states and Ishibashi wavefunctionals can be derived in an alternative
setting by using surgery techniques on
three-manifolds~\cite{FFFS1}. Let us compute
the one-point conformal correlation function on the disk
$\Sigma={\mathbb D}^2$ of a primary field of charge $\lambda$ with
boundary condition $\lambda'$. In the membrane picture, we compute
this correlator by inserting a vertical Wilson line of charge $\lambda$
through the connecting three-manifold $M_\Sigma={\mathbb D}^3$, linked
with the unknotted Wilson loop of charge $\lambda'$ representing
the boundary interaction as prescribed in the previous
section. This membrane state is depicted on the left in
Figure~\ref{1ptfndisk}. Since the corresponding Chern-Simons Hilbert
space is one-dimensional, this state is proportional to the
closed string state given by the Ishibashi boundary state on the
two-sphere $\sphere^2$. This basis vector for the physical Hilbert
space is represented on the right in Figure~\ref{1ptfndisk}.

We can determine the coupling coefficients in this setting by gluing
another three-ball ${\mathbb D}^3$ to each of the balls in
Figure~\ref{1ptfndisk} along their common (but oppositely oriented)
$\sphere^2$ boundaries. The left-hand side then gives the expectation
value of the Hopf link in $\sphere^3$ with component charges $\lambda$
and $\lambda'$, while the right-hand side gives the gauge theory
correlator on $\sphere^3$ of a single unknot of charge
$\lambda$. Using the functorial property of the quantum gauge theory,
this results in the equality
\beq
\bigl\langle W^{~}_{{\rm
    Hopf}(\lambda,\lambda')}\bigr\rangle_{\sphere^3}=
\beta_{\lambda'}^{~~\lambda}\,\bigl\langle W^{~}_{{\rm
    unknot}(\lambda)}\bigr\rangle_{\sphere^3} \ ,
\label{Hopfunknotrel}\eeq
and from (\ref{SmatrixHopf}) the Cardy solution (\ref{betaCardysol})
again unambiguously follows. With the canonical normalization of
boundary conformal field theory~\cite{FFFS1}, this gives the usual
expression for the one-point correlation function of a bulk field on the disk
${\mathbb D}^2$ as ${\sf S}_{\lambda\lambda'}/\sqrt{{\sf
    S}_{\lambda0}}$ times the standard two-point conformal block on
the sphere $\sphere^2$. Furthermore, gauge invariance of the
bulk theory implies that the branes obtained in this way correspond to
{\it all} relevant branes of the topological membrane~\cite{bb}.

\subsection{The Annulus Amplitude\label{TAA}}

\begin{figure}[ht]
\centerline{\epsfxsize=4cm
\epsfbox{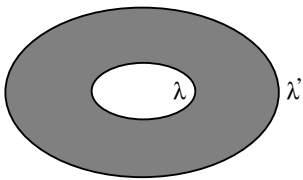}}
\caption{The annulus with prescribed conformal boundary conditions
  $\lambda$ and $\lambda'$. The inner radius is parametrized as
  $a=\e^{-t}$ where $t\in\real$ is the modulus of the annulus.}
\label{annulusampl}\end{figure}

As a final check of the consistency of our membrane identifications,
let us now look at the genus~$1$ annulus amplitude
(Figure~\ref{annulusampl}). By a direct calculation in the $\sf PCT$
orbifold of topologically massive gauge theory, it is given by the
Neumann amplitude~\cite{bb,CFKS1}
\beq
Z^{\sf PCT}_{\lambda\lambda'}(t)=
\sum_{\rho\in[\Lambda_{G_k}^*/\Lambda^{~}_{G_k}]^{~}_{\sf PCT}}
N_{\lambda\lambda'}^{~~~\rho}~\psi_\rho^{\sf PCT}(t)
\label{PCTampl}\eeq
where $N_{\lambda\lambda'}^{~~~\rho}$ are the fusion coefficients for
the Kac-Moody algebra based on the gauge group $G$ at level $k$. One
can compare this result with the Dirichlet cylinder amplitude computed
in a completely analogous way in the $\sf PT$ orbifold. The two
results are related after a modular transformation and a Poisson
resummation~\cite{CFKS1}, and immediately lead to the Verlinde formula
(\ref{Verlindeformula}) for the three-punctured sphere $\sphere^2$.

One can also arrive at this result via a surgery prescription~\cite{FFFS1}. The
connecting three-manifold in this case is a solid torus, with an
annulus Wilson graph representing the boundary interactions. Again,
one immediately has the expression (\ref{PCTampl}) for the annulus
amplitude. Let us now take the connected sum of the connecting
three-manifold with another solid torus. This produces the
Chern-Simons invariant of $\sphere^3$ with three unknots, and
functoriality then leads once again to the Verlinde formula for the
three-punctured two-sphere. This is completely analogous to the
original derivation~\cite{Witten1} of the Verlinde diagonalization formula in
terms of closed string amplitudes in three dimensions. Thus from the
point of view of topological membranes, the open string derivation of the
Verlinde formula~\cite{Cardy1} is completely equivalent to the closed
string derivation~\cite{Witten1}.

\section{Descendent States\label{DS}}
\setcounter{equation}{0}

As our final piece of evidence for the appearence of D-brane states,
we will now show how the dynamics of the topological membrane
naturally induces the correct tension of a D-brane. To derive the
tension formula, we shall need to further extend our construction of
vacuum Schr\"odinger wavefunctionals to describe excited membrane
eigenstates of the topologically massive gauge theory
Hamiltonian. At the same time, this will solve another problem of
topological membrane theory that we have not yet addressed, namely the
proper description of gauge invariant states in three dimensions
which correspond to string descendent fields.

\subsection{Landau Levels and Excited Membrane States\label{LLEMS}}

We will begin by constructing excited wavefunctionals of topologically
massive gauge theory in a fixed, non-trivial dilaton background of
the conformally coupled topologically massive gravity described in
Section~\ref{CCD}~\cite{bb}. For simplicity we will restrict our attention to a
$U(1)$ gauge group, or equivalently to the $c=1$ conformal field
theory of a single boson. Using the dilaton equation of motion arising
from variation of the action (\ref{SCTMGTdef}) with respect to the
scalar field $D$, the Hamiltonian of the gauge sector can be written
as
\beq
H_D=\frac12\,\int_\Sigma\,\left(\frac{e^2}{D^2}\,E^{0,1}\wedge E^{1,0}
+2\kappa~\dd{\rm vol}_\Sigma~V(D,\omega)\right) \ ,
\label{HDCTMGT}\eeq
where the dilaton potential $V(D,\omega)$ defined by
\beq
\dd{\rm vol}_\Sigma~V(D,\omega)=8D*_2\,\nabla^2D-D^2\,R^{(2)}(\omega)
\label{dilatonpot}\eeq
is essentially the three-dimensional energy density of the Liouville
field theory (\ref{defLiouville}). The dilaton background shifts the
ground state energy $\mathcal{E}=0$ to the non-vanishing value
\beq
\mathcal{E}_0=\kappa\,\bigl\langle V(D,\omega)\bigr\rangle \ .
\label{vacendilaton}\eeq

As observed previously, Hamiltonian quantization of topologically
massive gauge theory is equivalent to a field theoretic version of the
Landau problem. Using this observation, one can build higher membrane
states. We start from the vacuum wavefunctionals (\ref{Xilambdaidef}) which
are destroyed by the electric field annihilation operators as in
(\ref{vacstatecondn}),
\beq
E_z\Xi^{\{\lambda_i\}}\bigl[\{z_i\}\,;\,\mbf A\bigr]=0 \ .
\label{EzXidestroy}\eeq
A natural set of gauge-invariant excited states of the topological
membrane is then obtained via successive insertions of powers of the
electric field creation operators at the insertion points
$z_i\in\Sigma$ to give
\bea
\Psi_n^{\{\lambda_i\}}\bigl[\{z_i\}\,,\,\{n_i\}\,;\,\mbf A\bigr]
&=&\prod_{i=1}^n\,\frac1{\sqrt{n_i!}}\,\left(\frac{4\pi\ii}k\,
E_\bz(z_i)\right)^{n_i}\Xi^{\{\lambda_i\}}\bigl[\{z_i\}\,;\,\mbf A\bigr]
\nn\\&=&\e^{-\frac{|k|}{8\pi}\,\int_\Sigma A^{1,0}\wedge A^{0,1}}
\,\int\DD\phi~\e^{-\frac{|k|}{8\pi}\,\int_\Sigma(\,
\overline{\partial}\phi-2A^{1,0}\,)\wedge\partial\phi}\nn\\&&
\times\,\prod_{i=1}^n\,\frac1{\sqrt{n_i!}}\,\bigl(
A_z(z_i)-\partial_z\phi(z_i)\bigr)^{n_i}~\e^{\ii\lambda_i\,\phi(z_i)}
\label{excitedstates}\eea
with $n_i\in\nat_0$. As in (\ref{Xirepspaces}), these wavefunctionals are
regarded as operators on products of the corresponding representation
spaces,
\beq
\Psi_n^{\{\lambda_i\}}\bigl[\{z_i\}\,,\,\{n_i\}\,;\,\mbf A\bigr]
{}~\in~\mbf V \ ,
\label{excitedinV}\eeq
and they are eigenstates of the Hamiltonian (\ref{HDCTMGT}),
\beq
H_D\Psi_n^{\{\lambda_i\}}\bigl[\{z_i\}\,,\,\{n_i\}\,;\,\mbf A\bigr]
=\mathcal{E}_N\,
\Psi_n^{\{\lambda_i\}}\bigl[\{z_i\}\,,\,\{n_i\}\,;\,\mbf A\bigr] \ ,
\label{HDeigen}\eeq
where the excited state energies are those of Landau levels
\beq
\mathcal{E}_N=\mathcal{E}_0+\frac\mu{\left\langle D^4\right\rangle}
\,N
\label{ENLandau}\eeq
with
\beq
N=\sum_{i=1}^nn_i \ .
\label{levelnumber}\eeq

The wavefunctionals (\ref{excitedstates}) evidently describe gauge
invariant membrane excitations that correspond to string descendent
states at level $N$ given by (\ref{levelnumber}). They correspond to
$n$ gauge invariant combinations of external charged particles and
photons situated at the points $z_i\in\Sigma$. Strictly speaking, the
infrared limit $\mu\to\infty$ projects these states onto the lowest
Landau level $N=0$, i.e. $n_i=0~~\forall i=1,\dots,n$, but we shall
soon describe how appropriate field theoretic renormalizations
can be employed such that the higher Landau levels contribute to
quantities in the ground state of the topologically massive gauge
theory. In this way the topological photon mass (\ref{topmassTMGT})
will naturally set the mass scales for both perturbative and
non-perturbative states of the induced string theory.

\subsection{D-Brane Tension\label{D-BT}}

To derive a formula for the tension of a D-brane in the membrane
formalism, we will use the observation that the tension in open string
theory can be computed as the appropriate regulated dimension of the
conformal field theory state space corresponding to the one-graviton
vertex operators~\cite{HKMS1}. In three dimensional language, this
means that we need to compute the dimension of the Hilbert space
$\mathcal{H}_k(\lambda)$ spanned by the wavefunctionals
(\ref{excitedstates}) for $n=1$ and for a fixed $U(1)$ charge
$\lambda$, incorporating all Landau levels $N\geq0$. Of course, this
Hilbert space is infinite dimensional, but we can compute its
dimension as in (\ref{thermalpartfn},\ref{dimCSHilbert}) by treating
the membrane size $\beta$ as a regulator, performing an appropriate
renormalization, and then taking the high-temperature limit
$\beta\to0$. The calculation is much different than the derivation of
the Verlinde formula described in Section~\ref{VF}, because the
Hamiltonian of topologically massive gauge theory in higher Landau
levels does not vanish and the finite temperature amplitudes now
depend explicitly on both $\beta$ and the insertion points
$z_i\in\Sigma$.

However, the limit $\beta\to0$ shrinks the size of the membrane and so
does not properly induce string dynamics. Instead, one should
formulate the theory in the {\it dual} finite temperature formalism by
compactifying the Euclidean time direction on a circle of
circumference $\tilde\beta=1/\mu^2\beta$, and then take the equivalent
limit $\tilde\beta\to\infty$ which decompactifies the membrane~\cite{bb}. The
regulated dimension of the physical state space then computes a
modified version of the Verlinde formula defined by
\beq
{\rm reg}\,\dim\mathcal{H}_k(\lambda)=\lim_{\tilde\beta\to\infty}\,
\Tr^{~}_{\mathcal{H}_k(\lambda)}\left(\e^{-\tilde\beta\,H_D}
\right) \ .
\label{regdimdef}\eeq
The trace in (\ref{regdimdef}) can be represented as a membrane inner
product of $n=1$ wavefunctionals (\ref{excitedstates}) analogous to
those used before, which now however requires regularization by an
ultraviolet cutoff $\Lambda$. In the limit $\tilde\beta\to\infty$,
$\Lambda\to\infty$ with $\ln(\Lambda)/\tilde\beta$ held fixed, the
topological graviton mass undergoes a finite renormalization and only
the lowest Landau level $N=0$ contributes to the
amplitude. Corresponding to the $\sf PT$ orbifold of the topological
membrane, this calculation thereby results in the expression~\cite{bb}
\beq
{\rm reg}\,\dim\mathcal{H}_k(\lambda)=\left\langle\left.
\Psi_1^\lambda(z)\,\right|\,\Psi_1^\lambda(z)\right\rangle_{\rm ren}
=\frac{|k|}{4\pi} \ .
\label{regdimPTorb}\eeq

By using the identification (\ref{kRrel}), the dimension mass
formula~\cite{HKMS1}
\beq
\mathcal{M}^2=\alpha'\,(g_s)^{\chi(\Sigma)}\,\sqrt{
{\rm reg}\,\dim\mathcal{H}_k(\lambda)}
\label{dimmass}\eeq
agrees exactly with the formula for the tension of a D-brane wrapping
the target space circle $\sphere^1$ of radius $R$, with $\chi(\Sigma)$
the Euler characteristic of the Riemann surface $\Sigma$. Because of
bulk charge conservation, this result is independent of
$\lambda$. It also agrees with a direct calculation of
orbifold inner products that yield the correct Born-Infeld effective
action for the target space string dynamics within the membrane
formalism~\cite{bb}.

\section*{Acknowledgments}

The author thanks M.~Asorey, S.~Ganguli and D.~Polyakov for
interesting discussions and correspondence. This work was supported
in part by an Advanced Fellowship from the Particle Physics and
Astronomy Research Council~(U.K.).

\end{document}